\documentstyle[prd,aps,multicol,epsf,mathlett]{revtex}
\widetext
\renewcommand{\narrowtext}{\begin{multicols}{2} 
\global\columnwidth20.5pc}
\renewcommand{\widetext}{\end{multicols} \global\columnwidth42.5pc}
\def\Lrule{\vspace*{-0.2in}\noindent\vrule width3.5in height.2pt
  depth.2pt \vrule depth0em height1em}
\def\Rrule{\vspace{-0.1in}\hfill\vrule depth1em height0pt \vrule
  width3.5in height.2pt depth.2pt\vspace*{-0.1in}}
\def\bml{\begin{mathletter}}
\def\eml{\end{mathletter}}
\def\beq{\begin{equation}}
\def\eeq{\end{equation}}
\def\bea{\begin{eqnarray}}
\def\eea{\end{eqnarray}}
\def\ba{\begin{array}}
\def\ea{\end{array}}

\def\to{\rightarrow}
\def\a{\alpha}

\def\l{\lambda}
\def\m{\mu}

\def\z{\zeta}

\def\e{{\rm e}}
\def\tr{\,{\rm tr}\,}

\def\pf{\,{\rm Pf}\,}
\renewcommand{\theequation}{\arabic{section}.\arabic{equation}}
\begin{document}
\preprint{TIT-HEP-444}
\draft
\title{Massive random matrix ensembles 
at $\bbox{\beta}$ = 1 \& 4 : 
QCD in three dimensions}
\author{Taro Nagao}
\address{
Department of Physics, Graduate School of Science,
Osaka University,
Toyonaka, Osaka 560-0043, Japan
}
\author{Shinsuke M. Nishigaki}
\address{
Department of Physics, Faculty of Science,
Tokyo Institute of Technology,
Oh-okayama, Meguro, Tokyo 152-8551, Japan
}
\date{May 9, 2000}
\maketitle
\begin{abstract} 
The zero momentum sectors in effective theories of 
three dimensional QCD 
coupled to pseudoreal (two colors) and real (adjoint)
quarks in a classically parity-invariant manner
have alternative descriptions in terms of 
orthogonal and symplectic ensembles of random matrices.
Using this correspondence,
we compute 
finite-volume QCD partition functions and
correlation functions of Dirac operator eigenvalues
in a presence of finite quark masses of the order of 
the smallest Dirac eigenvalue.
These novel correlation functions,
expressed in terms of quaternion determinants,
are reduced to conventional results
for the Gaussian ensembles
in the quenched limit.
\end{abstract}
\renewcommand{\thefootnote}{\fnsymbol{footnote}}
\setcounter{footnote}{0}
\narrowtext
\section{introduction}
Spontaneous breaking of global symmetries has long been 
a subject of an extensive study
for quantum field theories in various dimensions.
Besides well-known cases such as
2D Gross-Neveu and Schwinger models and 4D QCD 
that exhibit breakdown of discrete or continuous 
chiral symmetry,
it has been suspected that 3D QED and QCD may also undergo
spontaneous breaking of flavor symmetry
\cite{Cor,JT,Red,Pis}.
For such odd dimensional field theories, 
there is no comprehensive theorem 
that predicts the surviving part of the flavor symmetry
as in the case of even dimensions \cite{VW},
as the notion of fermion
chirality is absent.
However, if there are even $N_f=2n$ number of 
massless 2-component complex fermions ${\psi}_j$, 
one can group them 
appropriately into 4-component complex fermions ${\Psi}_j$
\cite{JT,Red,Pis}:
\beq
{\cal L}=
\sum_{j=1}^{2n} \bar{\psi}_j D_\mu \sigma_\mu {\psi}_j
=
\sum_{j=1}^{n} \bar{\Psi}_j D_\mu \gamma_\mu {\Psi}_j,
\eeq
where $\gamma_\mu=\sigma_\mu \otimes \sigma_3 \ (\mu=1,2,3)$
in Euclidean 3D space.
Then one can define a {\em quasi}-chirality
according to two Hermitian matrices,
e.g. 
$\gamma_4=\openone \otimes \sigma_1$ and 
$\gamma_5=\openone \otimes \sigma_2$,
that anticommute with the Dirac operator
$D_\mu \gamma_\mu$ and with each other.
Generators of the flavor $U(2n)$ group are
rearranged into that of $U(n)$ group times
$\{ 1, \gamma_4, \gamma_5 , \gamma_S=\openone \otimes \sigma_3 \}$
in the 4-component notation.
In order to predict spontaneous flavor symmetry breaking pattern
along the same line as in 4D QCD,
one introduces a small symmetry-breaking mass term 
that is parity-invariant, i.e.,
$\sum_{j=1}^{n} m_j \bar{\Psi}_j {\Psi}_j$ 
but not 
$\sum_{j=1}^{n} m_j \bar{\Psi}_j \gamma_S{\Psi}_j$.
In the 2-component notation it leads to include
masses
$\{m\}=(m_1,\ldots,m_n,-m_1,\ldots,-m_n)$.
As the fermion determinant 
\beq
\prod_{j=1}^n \det(iD_\mu \gamma_\mu+i m_j)=
 \prod_{j=1}^n \det(-(D_\mu \sigma_\mu)^2+m_j^2)
\eeq
is positive definite under this restriction,
one can appeal to the Vafa-Witten theorem \cite{VW}
and predict that
if the flavor symmetry is spontaneously broken,
the absolute values of the fermion condensate
$\langle\bar{\psi}_j  {\psi}_j \rangle$ are equal for all 
$j=1,\ldots,2n$
and their signs are the same as those of respective masses.
That is, the continuous part of the global symmetry group is broken 
according to \cite{Pis,VZ3D}
\beq
U(2n) \rightarrow U(n)\times U(n)
\eeq
by the order parameter
\beq
\Sigma=\frac{1}{2n}\sum_{j=1}^{2n}
| \langle\bar{\psi}_j  {\psi}_j \rangle |,
\eeq
while the discrete $\bbox{Z}_2$ group
(a product of the parity and the exchange of fields 
$\psi_j\leftrightarrow \psi_{n+j}$)
remains unbroken.
The formation of the quark condensate was indeed observed
in Monte-Carlo simulations on a lattice \cite{DHKM}.
This peculiar pattern of flavor symmetry breaking can
also be predicted for 3D large-$N_c$ QCD
by the Coleman-Witten argument \cite{CW},
as remarked in Ref.\cite{VZ3D}.
One can repeat the above argument valid for 
complex (fundamental representation of $SU(N_c\geq 3)$ gauge group)
fermions towards the cases with
even flavors of  
pseudoreal (fundamental representation of $SU(2)$ gauge group) and 
of real (adjoint representation of $SU(N_c)$ gauge group) fermions.
We assign Dyson indices $\beta=2,1,4$, respectively, to these three cases,
according to the anti-unitary symmetries of
the associated Dirac operators \cite{V3foldway}.
Then the continuous parts of the global symmetry groups
are predicted to be broken down as \cite{Mag}
\bml
\bea
Sp(2n)&\rightarrow &Sp(n)\times Sp(n)\, \ \ \ \ \ (\beta=1) ,\\
SO(2n)&\rightarrow &SO(n)\times SO(n)\ \ \ \ (\beta=4) .
\eea
\eml
These symmetry breaking patterns determine the forms of 
the low-energy effective Lagrangian of associated
Nambu-Goldstone bosons:
\beq
{\cal L}_{\rm eff}=
\frac{F^2}{4} \tr \partial_\mu U \partial_\mu U^\dagger
-\Sigma \tr \bbox{m} U \Gamma U^\dagger + \cdots ,
\label{nlsm}
\eeq
where $U({\bf x})$ takes its value in the coset manifolds$^\dagger$
\beq
{\cal M}_F=
{\rm AIII}_{n,n},\ 
{\rm CII}_{n,n},\ 
{\rm BDI}_{n,n},
\eeq
respectively for $\beta=2, 1, 4$.
The mass and chiral matrices in the above are defined as
\bea
&& \bbox{m}=
\left\{
\ba{ll}
{\rm diag}(m_1,\ldots,m_n)\otimes \sigma_3 
& \ (\beta=2,4)\\
{\rm diag}(m_1,\ldots,m_n)\otimes \sigma_3 \otimes i\sigma_2
&\ (\beta=1)
\ea
\right. , \\
&& \Gamma=
\left\{
\ba{ll}
\openone_n \otimes \sigma_3 
&\ (\beta=2,4)\\
\openone_n\otimes \sigma_3 \otimes i\sigma_2
&\ (\beta=1)
\ea
\right. .
\eea
Given non-linear $\sigma$ models (NL$\sigma$Ms) in 4D, 
Leutwyler and Smilga
\cite{LS} proposed to extract out of them nonperturbative, exact
information on the spectra of Dirac operators 
by imposing a constraint on the parameters:
the linear dimension $L$ of the system
be much smaller than the Compton length of Nambu-Goldstone bosons
$\sim F/\sqrt{m\Sigma}$.
In this `ergodic' regime where
the zero momentum mode of $U({\bf x})$ dominates, the effective 
finite-volume partition function simplifies
into a finite-dimensional integral
and its small-$m$ expansion yields a sequence of
spectral sum rules.
Verbaarschot and collaborators
\cite{ShV,VZ} (see Ref.\cite{Vrev} for a
exhaustive list of references)
made an important observation along this line,
that the 0D finite-volume partition functions could as well be
derived from models much simpler than QCD,
random matrix ensembles (RMEs).
In our 3D QCD  context \cite{VZ3D,Mag}, it means that
in the limit
\beq
L\to \infty,\ \ \ m_i \to 0, \ \ \ \mu_i\equiv L^3 \Sigma m_i
\mbox{ : fixed},
\label{LSlim}
\eeq
the finite-volume partition functions
\bea
{\cal Z}(\{\m\})&=&\int_{{\cal M}_F}\!\!\!dU\,
\exp\left(\tr \bbox{\mu} U \Gamma U^\dagger\right), 
\label{Zfv}\\
\bbox{\mu}&\equiv& L^3 \Sigma \bbox{m}, \nonumber
\eea
have alternative representation in terms of
large-$N$ RMEs
\beq
{Z}(\{m\})=
\int_{\cal D} dH \e^{-\beta \tr V(H^2)}
\prod_{i=1}^n
\det 
\left(H^2+ m_j^2\right),
\label{ZRME}
\eeq
where the integral domains ${\cal D}$ are sets of $N\times N$
complex hermitian ($u(N)=T({\rm A}_{N})$), 
real symmetric ($o(N)=T({\rm AI}_{N})$), 
and quaternion selfdual ($sp(N)=T({\rm AII}_{N})$) matrices $H$ 
for $\beta=2,1,4$, respectively.
The determinant in the $\beta=4$ case is understood 
as a quaternion determinant (Tdet).
Their proofs consist of the `color-flavor' (or Hubbard-Stratonovich)
transformation
\cite{VWZ,Zircf} that converts the integration variables into
matrices with small ($n\times n$) dimensions,
and the saddle point method under which
\beq
N\to \infty,\ \ \ m_i \to 0, \ \ \ \mu_i\equiv\pi \bar{\rho}(0) m_i
\mbox{ : fixed}.
\label{microlim_m}
\eeq
Here $\bar{\rho}(0)$ stands for the large-$N$ 
spectral density of the random matrix
$H$:
\beq
\bar{\rho}(x)=\lim_{N\to\infty}
\langle \tr \delta(x-H) \rangle,
\eeq
at the spectral origin.
The RMEs (\ref{ZRME}) are motivated by
the microscopic theories (3D Euclidean QCD) on a lattice, 
with a crude simplification of replacing
matrix elements of the Hermitian Dirac operator
$i\,/\!\!\!\!D= (i\partial_\mu +  A_\mu)\sigma_\mu$ 
by random numbers $H_{jk}$ distributed according to 
the weight $\e^{-\beta \tr V(H^2)}$.
Under this correspondence, the microscopic limit (\ref{microlim_m})
is equivalent to Leutwyler-Smilga limit (\ref{LSlim}),
since the size $N$ of the matrix $H$ is interpreted as
the number of sites $L^3$ of the lattice on which QCD is discretized,
and the Dirac spectral density at zero virtuality $\bar{\rho}(0)$
is related to the quark condensate by the Banks-Casher relation
$\Sigma=\pi\bar{\rho}(0)/L^3$ \cite{BC}.

On the other hand, the situation is far more subtle in the case with 
an odd 
number ($N_f=2n+1$) of 2-component fermion flavors \cite{Red,VZ3D}.
In the massless case, the fermion determinant
$\det^{2n+1} (i D_\mu \sigma_\mu)$ is not positive definite,
and its phase is gauge dependent.
This dependence can be compensated by the Chern-Simons term
that is yielded by a gauge-invariant regularization, but
this anomalous term explicitly breaks the parity \cite{Red}.
Therefore, even in a presence of small masses
$\{m\}=(m_1, \ldots, m_n, -m_1, \ldots, -m_n, 0)$
that respects the $\bbox{Z}_2$ invariance
(combining parity and flavor exchange) 
classically,
one cannot appeal to the previous argument
to derive low-energy effective Lagrangians.
With this situation in mind, 
we nevertheless adopt a pattern of 
spontaneous flavor symmetry breaking
proposed by Verbaarschot and Zahed \cite{VZ3D} for $\beta=2$
and its generalizations to $\beta=1$ and 4:
\bml
\bea
 U(2n+1) &\rightarrow& U(n) \times U(n+1) \ \ \ \ \ \ \ \ (\beta=2),\\
Sp(2n+1) &\rightarrow& Sp(n)\times Sp(n+1)\ \ \ \ \ \ (\beta=1),\\
SO(2n+1) &\rightarrow& SO(n)\times SO(n+1)\ \ \ \ (\beta=4),
\eea
\eml
leading to NL$\sigma$Ms of 
Nambu-Goldstone fields over the coset manifolds
\beq
{\cal M}_F=
{\rm AIII}_{n,n+1},\ 
{\rm CII}_{n,n+1},\ 
{\rm BDI}_{n,n+1},
\eeq
respectively.
Then, by the same token as in the case of even $N_f$,
one can write down corresponding RMEs \cite{Chr}:
\beq
{Z} (\{m\})=
\int_{\cal D} dH \e^{-\beta \tr V(H^2)}
\det H
\prod_{i=1}^n
\det 
\left(H^2+ m_j^2\right) ,
\label{ZRMEodd}
\eeq
which are equivalent, in the limit (\ref{microlim_m}), to
the `finite-volume partition functions' \cite{VZ3D}
\bea
{\cal Z}(\{\m\})&=&\int_{{\cal M}_F}\!\!\!\!dU\!
\left\{
\ba{ll}
\cosh \left(\tr \bbox{\mu} U \Gamma U^\dagger\right) &
(N : {\rm even}) \\
\sinh \left(\tr \bbox{\mu} U \Gamma U^\dagger\right) &
(N : {\rm odd})
\ea
\right. \!\!\!\!,
\label{Zfvodd}\\
\Gamma&=&
\left\{
\ba{ll}
{\rm diag}(\openone_n , -\openone_{n+1})
&\  (\beta=2,4)\\
{\rm diag}(\openone_n , -\openone_{n+1})\otimes i\sigma_2
&\ (\beta=1)
\ea
\right. .
\eea
As the Chern-Simons term cannot be incorporated within
these RMEs, their physical relevance is unclear \cite{VZ3D}.
An immediate problem is that 
if the rank $N$ of the matrix $H$ is odd,
the partition function (\ref{ZRMEodd}) 
(or (\ref{Zfvodd})) is zero, and
the (unnormalized) correlation functions are odd under 
a simultaneous change of signs of the arguments,
which are unacceptable as physical observables.  
The above relationships between RMEs and NL$\sigma$Ms for 3D 
QCD consist, together with its counterpart for 4D QCD
(${\cal M}_F={\rm A}_{n}, {\rm AII}_{n}, {\rm AI}_{n}$ and
 ${\cal D}=T({\rm AIII}_{N,N'}), T({\rm BDI}_{N,N'}), 
 T({\rm CII}_{N,N'})$ for 
 $\beta=2, 1, 4$, respectively),
a part of Zirnbauer's
complete classification scheme of RMEs in terms of
Riemannian symmetric spaces \cite{Zir}.

These RMEs are technically suited for the computation
of correlations of eigenvalues $\{x\}$ of the Dirac operator 
$i\, /\!\!\!\!D \sim H$ in the microscopic asymptotic limit
where the energy eigenvalues are scaled 
as the quark masses are in Eq.(\ref{microlim_m}),
\beq
N\to\infty,\ \ x \to 0,\ \ 
\lambda= \pi \bar{\rho}(0)x : \mbox{fixed}.
\label{microlim_x}
\eeq
For the chiral RMEs describing 4D QCD in 
the low-energy ergodic regime,
such Dirac spectral correlators have been
computed previously for the massless 
\cite{VZ,NS,For,AST,NF95,NF98}
and recently for the massive cases 
\cite{JNZ,DN98a,WGW,NDW,NN1,NN2,AK} with all three
values of $\beta$.
On the other hand, for the (non-chiral) RMEs describing 3D QCD,
they have been analytically computed solely for the
unitary ($\beta=2$) ensemble, 
in the massless case \cite{NS,VZ3D},
as well as in the massive case \cite{DN98b,Chr}.
For other values of $\beta$, 
a numerical work based on a finite-$N$ formula
for the correlation functions recently appeared
only in the massless case \cite{HN}.
The subject of this Article is to complete this program
by analytically
computing the partition and correlation functions
for the orthogonal ($\beta=1$) and symplectic ($\beta=4$) ensembles
in a presence of finite scaled mass parameters.
We employ a slightly modified version of 
the method used in our previous articles
\cite{NN1,NN2}.

We finally remark on the universality issue.
It was noticed by
\c{S}ener and Verbaarschot \cite{SV} (see also Ref.\cite{KV})
and proved by Widom \cite{Wid} 
that the diagonal
element $S(x,y)$ of the quaternion kernel
for an orthogonal or symplectic ensemble, and accordingly 
all spectral correlation functions thereof, can be 
constructed from the scalar kernel $K(x,y)$ for a unitary ensemble
sharing the same weight function:
\bml
\bea
S^T&=&\Bigl(I-(I-K)\varepsilon K D\Bigr)^{-1} K 
\ \ \ \ \ (\beta=1), \\
S&=&\Bigl(I-(I-K)D K \varepsilon \Bigr)^{-1} K 
\ \ \ \ \ (\beta=4), 
\eea
\label{1.21}
\eml
\noindent
where $I$, $D$, $\varepsilon$, $S$, $K$ 
stand for integral operators with
convolution kernels $\delta(x-y)$,
$\delta'(x-y)$,  $\slantfrac{1}{2} {\rm sgn}(x-y)$, $S(x,y)$,
$K(x,y)$, respectively,
${}^T$ and ${}^{-1}$ stand for transpose and inverse
operators.
Since the scalar kernel in the asymptotic limit
(\ref{microlim_x}), (\ref{microlim_m})
is insensitive to the details of the potential $V(x^2)$
either in the absence \cite{BZ,ADMN} or
in the presence of finite and nonzero $\mu$'s \cite{DN98b,Chr}, 
the universality of correlation functions for 
orthogonal and symplectic ensembles are
automatically guaranteed${}^\ddagger$.
Therefore it suffices for us to concentrate onto Gaussian
ensembles, $V(x^2)= x^2/2$.
This choice leads to Wigner's semi-circle law
\beq
\bar{\rho}(x)=\frac{1}{\pi}\sqrt{2N-x^2}.
\eeq
\setcounter{equation}{0}
\widetext
\Lrule
\section{orthogonal ensemble}
For $\beta=1$, 
we treat the following three cases separately:
\begin{eqnarray*}
{\bf A}&:& \{m\}=(m_1,\ldots,m_n,-m_1,\ldots,-m_n),\\
{\bf B}&:& \{m\}=(m_1,\ldots,m_n,-m_1,\ldots,-m_n,0),\ {\rm even}\ N,\\
{\bf C}&:& \{m\}=(m_1,\ldots,m_n,-m_1,\ldots,-m_n,0),\ {\rm odd}\ N.
\end{eqnarray*}
\subsection{even ${\bf N}_{\bf f}$}
We first consider the case with $N_f\equiv 2n$
flavors and
$\{m\}=(m_1,\ldots,m_n,-m_1,\ldots,-m_n)$.

We express the partition function (\ref{ZRME})
of the RME in terms of eigenvalues $\{x_j\}$ 
of $H$ (up to a constant independent of $m$): 
\beq
{Z} (\{ m \})=\frac{1}{N!}
\int_{-\infty}^\infty 
\!\!\!\!\cdots \int_{-\infty}^\infty 
\prod_{j=1}^N dx_j
\prod_{j=1}^N
\biggl(
{\rm e}^{-x_j^2/2}  
\prod_{k=1}^n (x_j^2+m_k^2)
\biggr)
\prod_{j>k}^N |x_j-x_k| .
\label{2.3}
\eeq
The $p$-level correlation function of
the matrix $H$ is defined as
\bea
\rho(x_1,\ldots,x_p;\{m\})
&=&
\langle \prod_{j=1}^p {\rm tr}\,\delta(x_j-H) \rangle
\nonumber\\
&=&
\frac{\Xi_p (x_1,\ldots,x_p;\{ m \})}{\Xi_0 (\{ m \})},
\label{rho_p}\\
\Xi_p (x_1,\ldots,x_p;\{ m \})&=&
\frac{1}{(N-p)!}
\int_{-\infty}^\infty \!\!\!\!\cdots \int_{-\infty}^\infty 
\prod_{j=p+1}^N dx_j 
\prod_{j=1}^N
\biggl(
{\rm e}^{-x_j^2/2}  
\prod_{k=1}^n (x_j^2+m_k^2)
\biggr)
\prod_{j>k}^N |x_j-x_k|
\label{Xi_p}
\eea
$(\Xi_0=Z)$.
We define new variables $z_j$ as 
\begin{eqnarray}
&&z_{2 j - 1}  =  i m_j \ \ \ (j=1,\ldots,{n}), \nonumber \\ 
&&z_{2 j}  =  - i m_j \ \ \ \ (j=1,\ldots,{n}),  \\ 
&&z_{2 {n} + j}  =  x_j \ \ \ \ \ (j= 1,\ldots,p).\nonumber
\end{eqnarray}
Then the multiple integral (\ref{Xi_p}) is expressed as
\bea
\Xi_p(z_1,\ldots,z_{2n+p})& = &
\frac{1}{\prod_{j=1}^{2 {n}} \sqrt{w(z_j)}
\prod_{j > k}^{2 {n}} (z_j - z_k )}
\nonumber\\
&&\times
\frac{1}{(N-p)!}
\int_{-\infty}^\infty\!\!\!\!\cdots\int_{-\infty}^\infty 
\!\!\prod_{j=2n+p+1}^{2n+N} dz_j 
\prod_{j=1}^{2n + N}\sqrt{w(z_j)} 
\prod_{j > k}^{2n + N} ( z_j - z_k)
\prod_{j > k>2n}^{2n + N} {\rm sgn}( z_j - z_k),
\label{Xi_p_z}
\eea
where 
$w(z) = {\rm e}^{- z^2}$.
Eq.(\ref{Xi_p_z}) resembles an 
$(2n + p)$-level correlation function of 
the conventional Gaussian ensemble 
with $2n + N$ levels. 
However, conventionally the levels 
$z_1,\ldots,z_{2n+p}$ are all real, 
while in the present case some of 
them ($z_1,\ldots,z_{2n}$) are pure imaginary. 
We carefully incorporate 
this fact into the following evaluation.

Let us denote 
the integrand in Eq.(\ref{Xi_p_z}) as
\beq
p(z_1,\ldots,z_{2 {n} + N}) =
\prod_{j=1}^{2 {n} + N} \sqrt{w(z_j)} 
\prod_{j > k}^{2 {n} + N} (z_j - z_k) 
\prod_{j > k > 2 {n}}^{2 {n} + N} {\rm sgn}(z_j - z_k).
\label{2.7}
\eeq
For $N$ even,
an identity
\begin{equation}
\prod_{j>k}^{2 {n} + N} {\rm sgn}(z_j - z_k) = 
{\rm Pf}[{\rm sgn}(z_k - z_j)]_{j,k = 1,\ldots,2 {n} + N}
\end{equation}
holds for {\em real} $z_1, \ldots, z_{2 {n} + N}$. 
By taking the limit 
$z_{1} < z_2 < \ldots < z_{2 {n}} \rightarrow - \infty$, 
we find another identity
\begin{equation}
\prod_{j>k>2 {n}}^{2 {n} + N} {\rm sgn}(z_j - z_k) = 
{\rm Pf}[F_{jk}]_{j,k = 1,\ldots,2 {n} + N},
\label{peven}
\end{equation}
where
\begin{equation}
F_{jk} = \left\{ \begin{array}{ll} 
{\rm sgn}(k - j) & \ \ 
(j,k = 1,\ldots,2 {n}) \\  
1& \ \ 
(j=1,\ldots,2 {n};\ k = 2 {n} + 1,\ldots,2 {n} + N)
\\
-1 & \ \ 
(j = 2 {n} + 1,\ldots,2 {n} + N ;\  k=1,\ldots,2 {n})
\\
{\rm sgn}(z_k - z_j) & \ \ 
(j,k = 2 {n} + 1,\ldots,2 {n} + N)
\end{array} \right.  .
\end{equation}
Substitution of Eq.(\ref{peven}) into Eq.(\ref{2.7}) yields 
\beq
p(z_1,\ldots,z_{2 {n} + N}) = 
 \prod_{j=1}^{2 {n}+N} 
\sqrt{w(z_j)} \prod_{j>k}^{2 {n}+N} (z_j - z_k) 
{\rm Pf}[F_{jk}]_{j,k = 1,\ldots,2 {n}+N} .
\eeq
For $N$ odd, we similarly obtain 
\beq
p(z_1,\ldots,z_{2 {n} + N}) = \prod_{j=1}^{2 {n} + N} 
\sqrt{w(z_j)} \prod_{j>k}^{2 {n}+N} (z_j - z_k) 
{\rm Pf} \left[ \begin{array}{ll}[F_{jk}]_{ j,k = 1,\ldots,2 {n}+ N}
 & [g_j]_{j=1,\ldots,2 {n}+N} \\
\left[-g_k \right]_{k=1,\ldots,2 {n}+N} & 0 \end{array} \right], 
\eeq
with
$g_j = g_k = 1$ ($j,k = 1,\ldots,2 {n}+N$).
The Pfaffians in the above can be represented as
quaternion determinants \cite{NF95,NF98,DysonQ,Meh,MehtaQ,MM,NF99,FNH}.
In doing so, we introduce monic skew-orthogonal polynomials
$R_j(z)= z^j + \cdots$,
which satisfy the skew-orthogonality relation:
\bea
&&\langle R_{2j}, R_{2k+1} \rangle_R =
 - \langle R_{2k+1}, R_{2j}\rangle_R = r_j \delta_{jk},\\
&&\langle R_{2j}, R_{2k} \rangle_R =
 \langle R_{2j+1}, R_{2k+1}\rangle_R = 0,\nonumber
\eea
where
\beq
\langle f, g \rangle_R
=
\int_{-\infty}^{\infty} dz 
\sqrt{w(z)} 
g(z) 
\int_{-\infty}^{z} {dz'} 
\sqrt{w(z')}
f(z') 
 - (f\leftrightarrow g).
\eeq
Explicit forms 
for the skew-orthogonal polynomials
and their norms associated with the Gaussian weight $w(z)$
are known \cite{Meh}:
\bea
R_{2j}(z)&=&\frac{1}{2^{2j}} H_{2j}(z),
\nonumber\\
R_{2j+1}(z)&=&\frac{1}{2^{2j+1}} \Bigl(
H_{2j+1}(z)-
H'_{2j}(z)
 \Bigr),
\label{Rn}\\
r_j&=&2^{-2j+1} (2j)! \sqrt{\pi} ,
\nonumber
\eea
in terms of the Hermite polynomials
\beq
H_{j}(z)=(-1)^j \e^{z^2} \frac{d^j}{dz^j} \e^{-z^2}.
\eeq

Now we present the following theorems:
\par
\bigskip
\noindent
{\em Theorem 1}
\par
\bigskip
\noindent
For even $N$, we can rewrite $p(z_1,\ldots,z_{2 {n}+N})$ as 
\beq
p(z_1,\ldots,z_{2{n} + N}) =
\Bigl( \prod_{j=0}^{{n}+N/2-1} r_{j}  \Bigl)
{\rm Tdet} [f_{jk}(z_j,z_k)]_{j,k = 1,\ldots,2 {n}+N}.
\eeq
The quaternion elements $f_{jk}(z_j,z_k)$ are represented as
\begin{equation}
f_{jk}(z_j,z_k)= \left[ \begin{array}{cc}
S(z_j,z_k) & I(z_j,z_k) \\ 
D(z_j,z_k) & S(z_k,z_j) \end{array} \right].
\eeq
The functions $S(z_j,z_k)$, $D(z_j,z_k)$ and $I(z_j,z_k)$
are given by
\bea
S(z_j,z_k) &=& \sum_{\ell=0}^{{n}+N/2-1}
\frac{
\Phi_{2\ell}(z_j) \Psi_{2\ell+1}(z_k) -
\Phi_{2\ell+1}(z_j) \Psi_{2\ell}(z_k)}{r_{\ell}},
\nonumber\\
D(z_j,z_k) &=& \sum_{\ell=0}^{{n}+N/2-1}
\frac{
\Psi_{2\ell}(z_j) \Psi_{2\ell+1}(z_k) -
\Psi_{2\ell+1}(z_j) \Psi_{2\ell}(z_k)}{r_{\ell}},
\label{SDI}\\
I(z_j,z_k) &=& -\sum_{\ell=0}^{{n}+N/2-1}
\frac{
\Phi_{2\ell}(z_j) \Phi_{2\ell+1}(z_k) - 
\Phi_{2\ell+1}(z_j) \Phi_{2\ell}(z_k)}{r_{\ell}} 
+F_{jk},
\nonumber
\eea
where
\bea
&&\Psi_j(z) = \sqrt{w(z)} R_j(z), 
\nonumber\\
&&\Phi_j(z_k) 
=
\left\{
\ba{lll}
\int_{-\infty}^{\infty}{dz} \sqrt{w(z)} R_j(z)\,{\rm sgn}(z_k-z)
& \ \ (k=2n+1,\ldots,2n+N)\\
-\int_{-\infty}^{\infty}{dz} \sqrt{w(z)} R_j(z) 
&\ \  (k: {\rm otherwise})
\ea
\right. .
\eea
\label{PsiPhi}
\par
\bigskip
\noindent
{\em Theorem 2}
\par
\bigskip
\noindent
For odd $N$, we have
\beq
p(z_1,\ldots,z_{2 {n} + N}) = 
\Bigl( \prod_{j=0}^{ {n}+[N/2]-1} r_{j} \Bigr) s_{2 {n}+N-1}\,
{\rm Tdet} 
[f_{jk}^{\rm odd}(z_j,z_k)]_{j,k = 1,\ldots,2 {n}+N}.  
\eeq
The quaternion elements are represented as
\begin{equation}
f_{jk}^{\rm odd}(z_j,z_k) = \left[ \begin{array}{cc} 
S^{\rm odd}(z_j,z_k) 
& I^{\rm odd}(z_j,z_k) \\ 
D^{\rm odd}(z_j,z_k) & S^{\rm odd}(z_k,z_j) \end{array} \right].
\eeq
and 
\beq
s_j=\int_{-\infty}^\infty \Psi_j(z)dz.
\label{sn}
\eeq
The functions $S^{\rm odd}$, $D^{\rm odd}$ and $I^{\rm odd}$ are 
given in terms of $S$, $D$ and $I$ defined
in Eq.(\ref{SDI}), according to
\begin{eqnarray}
{S^{\rm odd}}(z_j,z_k) &= & S(z_j,z_k) \Big|_\ast 
+ {{\Psi}_{2 {n} + N-1}(z_k) \over s_{2 {n} + N-1}},
\nonumber\\
{D^{\rm odd}}(z_j,z_k) & = & D(z_j,z_k) \Big|_\ast ,\\
{I^{\rm odd}}(z_j,z_k)  & = & I(z_j,z_k) \Big|_\ast + 
{{\Phi}_{2 {n} + N-1}(z_j) - {\Phi}_{2 {n} + N-1}(z_k) 
\over s_{2n+N-1}} .  
\nonumber
\end{eqnarray}
Here $\ast$ stands for a set of substitutions
\beq
{R}_j(z) \mapsto 
R_j(z)-\frac{s_j}{s_{2n+N-1}}R_{2n+N-1}(z)
\ \ \ \ \ (j=0,\ldots,2 {n} + N-2),
\eeq
associated with a change in the upper limit of the sum
\beq
{n} + \frac{N}{2} -1 \mapsto {n} + \left[ \frac{N}{2} \right]-1.
\eeq
\par
\bigskip
\noindent
{\em Theorem 3}
\par
\bigskip
\noindent
Let the quaternion elements $q_{jk}$ of a selfdual $n \times n$ matrix 
$Q_n$ 
depend on $n$ real or complex variables $z_1,\cdots,z_n$ as 
\begin{equation} 
q_{jk} = f_{jk}(z_j,z_k). 
\end{equation}
We assume that $f_{jk}(z_j,z_k)$ satisfies the following conditions.
\bml
\bea
&&\int f_{nn}(z_n,z_n) d\mu(z_n) = c_n, \\
&&
\int f_{jn}(z_j,z_n) f_{nk}(z_n,z_k) d\mu(z_n) 
= f_{jk}(z_j,z_k) +  
 \lambda f_{jk}(z_j,z_k) - f_{jk}(z_j,z_k) \lambda.
\eea
\eml
Here $d\mu(z)$ is a suitable measure, $c_n$ is a 
constant scalar, and $\lambda$ is a constant quaternion. Then we 
have
\begin{equation} 
\int {\rm Tdet}\, Q_n \, d\mu(z_n) = 
(c_n-n+1) {\rm Tdet}\, Q_{n-1},
\end{equation}
where $Q_{n-1}$ 
is the $(n-1) \times (n-1)$ matrix obtained by removing the row 
and the column which contain $z_n$.
\par
\bigskip
It is straightforward to show that 
the quaternion elements 
$f_{jk}(z_j,z_k)$ 
and $f_{jk}^{\rm odd}(z_j,z_k)$ in {\em Theorem 1} and {\em Theorem 2}, 
respectively, both satisfy the conditions imposed 
on $f_{jk}(z_j,z_k)$ in {\em Theorem 3}.
This means that we can inductively write
\beq
\Xi_p(z_1,\ldots,z_{2n+p}) =
\frac{\prod_{j=0}^{p+[N/2]-1} r_{j} }{\prod_{j=1}^{2n} 
\sqrt{w(z_j)} 
\prod_{j > k}^{2n} ( z_j - z_k )}\times
\left\{
\ba{lll}
           &{\rm Tdet}[f_{jk}(z_j,z_k)]_{j,k=1,\ldots,2n+p}
& (N : \mbox{even})\\
s_{2n+N-1}\!\! & 
{\rm Tdet}[{f_{jk}^{\rm odd}}(z_j,z_k)]_{j,k=1,\ldots,2n+p}
& (N: \mbox{odd})
\ea
\right.  .
\eeq
Since the final result in the asymptotic limit $N \rightarrow \infty$ 
should be insensitive to the parity of $N$, we consider only even 
$N$ henceforth. Then the $p$-level correlation function 
(\ref{rho_p})
is written as
\beq
\rho(x_1,\ldots,x_p;\{m\}) 
= \frac{\Xi_p(z_1,\ldots,z_{2n+p})}{\Xi_0(z_1,\ldots,z_{2n})}
=\frac{{\rm Tdet}[f_{jk}(z_j,z_k)]_{j,k=1,\ldots,2 {n}+p}}{{\rm 
Tdet}[f_{jk}(z_j,z_k)]_{j,k=1,\ldots,2 {n}}}.
\eeq

We introduce a set of notations 
\begin{eqnarray}
&&
S^{II}_{jk} = S(z_j,z_k) 
\ \ \ \ \ \ \ \ \ \ \ \ \ \ 
(j,k = 1,\ldots, 2 {n}), 
\nonumber\\ 
&& 
S^{IR}_{jk} = S(z_j,z_{2 {n} + k}) 
\ \ \ \ \ \ \ \ \ 
(j = 1,\ldots, 2 {n} ;\ k = 1,\ldots,p),
\nonumber\\
&&
S^{RI}_{jk} = S(z_{2 {n} + j},z_k) 
\ \ \ \ \ \ \ \ \ 
(j = 1,\ldots,p ;\ k= 1,\ldots, 2 {n}) ,
\\  
&&
S^{RR}_{jk} = S(z_{2 {n} + j},z_{2 {n} + k}) 
\ \ \ \ (j,k = 1,\ldots, p)\nonumber
\end{eqnarray}
and similarly for $D$ and $I$.
Using Dyson's identity (\ref{Dysoneq})
we can rewrite the correlation functions as 
\begin{eqnarray}
\rho(x_1,\ldots,x_p;\{m\}) &=&  
(-1)^{p(p-1) / 2} \frac{
{\rm Pf} \left[ \begin{array}{cccc}
-I^{II} & S^{II} & -I^{IR} & S^{IR} \\ 
-(S^{II})^{T} & D^{II} & -(S^{RI})^{T} & D^{IR} \\  
-I^{RI} & S^{RI} & -I^{RR} & S^{RR} \\ 
-(S^{IR})^{T} & D^{RI} & -(S^{RR})^{T} & D^{RR} \end{array} \right]  
}{{\rm Pf} \left[ \begin{array}{cc}
-I^{II} & S^{II} \\ 
-(S^{II})^{T} & D^{II} 
\end{array} \right]} 
\nonumber \\ 
&=&  (-1)^{p(p-1) / 2} \frac{
{\rm Pf} \left[ \begin{array}{ccc}
 D^{II} & -(S^{RI})^{T} & -(D^{RI})^T \\  
 S^{RI} & -I^{RR} & S^{RR} \\ 
 D^{RI} & -(S^{RR})^{T} & D^{RR} \end{array} \right]  
}{{\rm Pf} \left[ D^{II} \right]}.
\label{rho1}
\end{eqnarray}
In the last line we have exploited a Pfaffian identity
that holds for antisymmetric matrices $A$, $B$
of even ranks 
and a row vector $v$:
\beq
{\rm Pf}
\left[
\ba{c|c}
A &  
\ba{c}
v\\
\vdots\\
v
\ea
\\
\hline
-v^T \cdots -v^T & B
\ea
\right]
={\rm Pf}[A] \,{\rm Pf}[B].
\eeq

Now we proceed to evaluate the component functions
of the quaternion kernel $f_{jk}(z_j,z_k)$ in
the asymptotic limit 
(\ref{microlim_m}) and (\ref{microlim_x}), where
unfolded microscopic variables 
\begin{eqnarray}
\sqrt{2N}z_{2 j - 1}&\equiv&
\z_{2 j - 1}  =  i \m_j \ \ \ \ (j=1,\ldots,{n}), \nonumber \\ 
\sqrt{2N}z_{2 j }&\equiv&
\z_{2 j }  =  -i \m_j \ \ \ \ \ (j=1,\ldots,{n}), \\ 
\sqrt{2N}z_{2n+ j }&\equiv&
\l_j \ \ \ \ \ \ \ \ \ \ \ \ \ \ \ \ 
(j= 1,\ldots,p)\nonumber
\end{eqnarray}
are kept fixed.
We note that all elements of
the sub-matrices that appear in the second line of Eq.(\ref{rho1})
are expressed as (derivatives or integrals of)
an analytic function
\begin{eqnarray}
{\bar S}(z,z')  &=&  
\sum_{j=0}^{n+N/2-1} 
\frac{\bar{\Phi}_{2j}(z)\Psi_{2j+1}(z')-
\bar{\Phi}_{2j+1}(z)\Psi_{2j}(z')}{r_{j}} 
\nonumber \\ 
& = & \frac{{\rm e}^{-(z^2+z'{}^2)/2}}{2^{2n+N} \sqrt{\pi} 
\Gamma(2n+N)} \frac{H_{2n+N}(z) H_{2n+N-1}(z') 
- H_{2n+N-1}(z) 
H_{2n+N}(z')}{z-z'} 
\nonumber\\
&&+ \frac{{\rm e}^{-z'{}^2/2}}{2^{2n+N} \sqrt{\pi} 
\Gamma(2n+N)} H_{2n+N-1}(z') 
\int_0^z {\rm e}^{-u^2/2} H_{2n+N}(u) {d}u,
\label{Szz}
\end{eqnarray}
where
\begin{equation}
{\bar \Phi}_j(z) 
= \left\{ \int_{-\infty}^z - \int_z^{\infty} 
\right\} \Psi_j(u) {d}u,
\label{barPhi}
\end{equation}
with real and/or imaginary arguments:
\bml
\bea
D^{II}_{jk} &=& \frac{1}{2} \frac{\partial}{\partial z_j} 
{\bar S}(z_j,z_k), \\
S^{RI}_{jk} &=& {\bar S}(x_j,z_k), \\
D^{RI}_{jk} &=&
\frac{1}{2} 
\frac{\partial}{\partial x_j} 
{\bar S}(x_j,z_k) ,\\
S^{RR}_{jk} &=& {\bar S}(x_j,x_k), \\
D^{RR}_{jk} &=& \frac{1}{2} 
\frac{\partial}{\partial x_j} 
{\bar S}(x_j,x_k),\\
I^{RR}_{jk} &=& - 2 \int_{x_j}^{x_k} 
{\bar S}(x_j,x) {d}x 
- {\rm sgn}(x_j - x_k) .
\eea
\eml
In the second line of Eq.(\ref{Szz}),
we have singled out the unitary scalar kernel and
applied to it the Christoffel-Darboux formula.
Substituting asymptotic formulas for
the Hermite polynomials
\begin{eqnarray}
H_{2k}(z) & \sim & \frac{ (-1)^k 2^{2 k} k!}{\sqrt{\pi k}} 
\cos(2 \sqrt{k} z), \nonumber \\ 
H_{2k+1}(z) & \sim & \frac{ (-1)^k 2^{2 k+1} k!}{\sqrt{\pi}} 
\sin(2 \sqrt{k} z),
\label{Hermiteasy}
\end{eqnarray}
valid under $k\to \infty$, $z\to 0$, $\sqrt{k} z:$ fixed,
one can show that $\bar{S}(z,z')$ approaches
the sine kernel \cite{Meh}:
\beq
\frac{1}{\sqrt{2N}} 
{\bar S}(\frac{\z}{\sqrt{2N}},\frac{\z'}{\sqrt{2N}})
\sim
\frac{\sin(\z-\z')}{\pi(\z-\z')}\equiv 
K(\zeta-\zeta') .
\eeq
The Pfaffian elements in the asymptotic limit,
after taking into account an unfolding by the factor
$\sqrt{2N}$, are then expressed in terms of $K(\z)$:
\bml
\bea
{\bf D}^{II}_{jk} &\equiv&
\frac{1}{2N}D(z_j,z_k) 
\sim \frac{1}{2} K'(\z_j-\z_k) ,
\\
{\bf S}^{RI}_{jk}& \equiv&
\frac{1}{\sqrt{2N}} S(x_{j},z_k) 
\sim 
K(\l_{j}-\z_k) ,\\
{\bf D}^{RI}_{jk} 
&\equiv&
\frac{1}{2N}D(x_{j},z_k) 
\sim \frac{1}{2} 
K'(\l_{j}-\z_k),\\
{\bf S}^{RR}_{jk} &\equiv&
\frac{1}{\sqrt{2N}} 
S(x_{j},x_{k})  
\sim K(\l_{j}-\l_{k}),\\
{\bf D}^{RR}_{jk} &\equiv&
\frac{1}{2N} D(x_{j},x_{k}) 
\sim \frac{1}{2} 
K'(\l_{j}-\l_{k}),
\\
{\bf I}^{RR}_{jk} &\equiv&I(x_{j},x_{k}) 
\sim {2} \int_0^{\l_{j}-\l_{k}} 
\!\!\!\!\!\!\!\!\!\!\!\!\!
K(\l) 
{d}\l - {\rm sgn}(\l_{j} - \l_{k}).
\eea
\label{SDI1}
\eml
\noindent
These matrix elements constitute
the finite-volume partition function
\begin{eqnarray}
{\cal Z}(\{\m\})
&\equiv& 
\Xi_0(\{ \frac{\m}{\sqrt{2N}} \})
\nonumber\\
&=&{\rm const.}\frac{{\rm Pf}[{\bf D}^{II}]}{
\prod_{j=1}^n \m_j \prod_{j>k}^n (\m_j^2-\m_k^2)^2},
\label{calZ1}
\eea
and the scaled spectral correlation functions
\begin{eqnarray}
\rho_s(\l_1,\ldots,\l_p;\{ \m \})
&\equiv &
\Bigl(\frac{1}{\sqrt{2N}}\Bigr)^p
\rho(\frac{\l_1}{\sqrt{2N}},\ldots,\frac{\l_p}{\sqrt{2N}};
\{ \frac{\m}{\sqrt{2N}} \})
\nonumber \\ 
&=&  (-1)^{p(p-1) / 2} \frac{
{\rm Pf} \left[ \begin{array}{ccc}
 {\bf D}^{II} & -({\bf S}^{RI})^{T} & -({\bf D}^{RI})^{T} \\  
 {\bf S}^{RI} & -{\bf I}^{RR} & {\bf S}^{RR} \\ 
 {\bf D}^{RI} & -({\bf S}^{RR})^{T} & {\bf D}^{RR}
\end{array} \right]  
}{{\rm Pf} \left[ {\bf D}^{II} \right]}.  
\label{rhos1}
\end{eqnarray}
In the quenched limit $\mu_1,\ldots,\mu_n\to \infty$
when the ratio of two Pfaffians is replaced by a minor 
$\pf 
\left[{
\,-{\bf I}^{RR} \ \ \ \ {\bf S}^{RR}   \atop
 -({\bf S}^{RR})^{T}\ {\bf D}^{RR} }\right]
$,
the correlation functions approach
those of the Gaussian orthogonal ensemble \cite{Meh}.
By the same token, it satisfies a sequence
\beq
\rho_s(\{\l\};\mu_1,\ldots,\mu_n,-\m_1,\ldots,-\mu_n)
\stackrel{\mu_{n} \to \infty}{\longrightarrow}
\rho_s(\{\l\};\mu_1,\ldots,\mu_{n-1},-\m_1,\ldots,-\mu_{n-1})
\stackrel{\mu_{n-1} \to \infty}{\longrightarrow}
\cdots ,
\label{decouple}
\eeq
as each of the masses is decoupled by going to infinity.
To illustrate this decoupling, we exhibit in FIG.1 a plot of
the spectral density $\rho_s(\l;\m,-\m)$ 
($p=1, n=1$).
\subsection{odd ${\bf N}_{\bf f}$, even ${\bf N}$}
Next we consider the case with $N_f\equiv 2n+1$
flavors,
$\{m\}=(m_1,\ldots,m_n,-m_1,\ldots,-m_n,0)$, and with even $N$.
We express the partition function (\ref{ZRMEodd})
of the RME in terms of eigenvalues $\{x_j\}$ of $H$:
\begin{eqnarray}
&&{Z} (\{ m \})=\frac{1}{N!}
\int_{-\infty}^\infty 
\!\!\!\!\cdots \int_{-\infty}^\infty 
\prod_{j=1}^N dx_j
\prod_{j=1}^N
\biggl(
{\rm e}^{-x_j^2/2}  x_j
\prod_{k=1}^n (x_j^2+m_k^2)
\biggr)
\prod_{j>k}^N |x_j-x_k| .
\label{2.45}
\end{eqnarray}
The $p$-level correlation function of
the matrix $H$ is defined as
\bea
\rho(x_1,\ldots,x_p;\{m\})
&=&
\langle \prod_{j=1}^p {\rm tr}\,\delta(x_j-H) \rangle
\nonumber\\
&=&
\frac{\Xi_p (x_1,\ldots,x_p;\{ m \})}{\Xi_0 (\{ m \})},
\label{rho_pb}\\
\Xi_p (x_1,\ldots,x_p;\{ m \})&=&
\frac{1}{(N-p)!}
\int_{-\infty}^\infty \!\!\!\!\cdots \int_{-\infty}^\infty 
\prod_{j=p+1}^N dx_j
\prod_{j=1}^N
\biggl(
{\rm e}^{-x_j^2/2}  x_j
\prod_{k=1}^n (x_j^2+m_k^2)
\biggr)
\prod_{j>k}^N |x_j-x_k|
\label{Xi_pb}
\eea
$(\Xi_0=Z)$.
We define new variables $z_j$ as 
\begin{eqnarray}
&&z_{0}  =  0,\nonumber\\
&&z_{2 j - 1}  =  i m_j \ \ \ (j=1,\ldots,{n}), \nonumber \\ 
&&z_{2 j}  =  - i m_j \ \ \ \ (j=1,\ldots,{n}),  
\label{2.46}\\ 
&&z_{2 {n} + j}  =  x_j \ \ \ \ \ (j= 1,\ldots,p).\nonumber
\end{eqnarray}
Then the multiple integral (\ref{Xi_pb}) is expressed as
\bea
\Xi_p(z_0,\ldots,z_{2n+p})& = &
\frac{1}{\prod_{j=0}^{2 {n}} \sqrt{w(z_j)}
\prod_{j > k\geq 0}^{2 {n}} (z_j - z_k )} 
\nonumber\\
&&\times
\frac{1}{(N-p)!}
\int_{-\infty}^\infty\!\!\!\!\cdots\int_{-\infty}^\infty 
\prod_{j=2n+p+1}^{2n+N} dz_j 
\prod_{j=0}^{2n + N}\sqrt{w(z_j)} 
\prod_{j > k\geq 0}^{2n + N} ( z_j - z_k)
\prod_{j > k>2n}^{2n + N} {\rm sgn}( z_j - z_k),
\label{Xi_p_zB}
\eea
where $w(z) = {\rm e}^{- z^2}$.

Let us denote 
the integrand in Eq.(\ref{Xi_p_zB}) as
\beq
p(z_0,\ldots,z_{2 {n} + N}) =
\prod_{j=0}^{2 {n} + N} \sqrt{w(z_j)} 
\prod_{j > k\geq 0}^{2 {n} + N} (z_j - z_k) 
\prod_{j > k > 2 {n}}^{2 {n} + N} {\rm sgn}(z_j - z_k).
\label{2.47}
\eeq
An identity
\begin{equation}
\prod_{j>k\geq 0}^{2 {n} + N+1} {\rm sgn}(z_j - z_k) = 
{\rm Pf}[{\rm sgn}(z_k - z_j)]_{j,k = 0,\ldots,2 {n} + N+1}
\end{equation}
holds for {\em real} $z_0, \ldots, z_{2 {n} + N+1}$. 
By taking the limit 
$z_0 < z_1 < \ldots < z_{2 {n}} \rightarrow - \infty$
and $z_{2 {n}+N+1} \rightarrow + \infty$, 
we find another identity
\beq
\prod_{j>k>2n}^{2 {n} + N+1} {\rm sgn}(z_j - z_k)
=
{\rm Pf} 
\left[ \begin{array}{ll}[F_{jk}]_{ j,k = 0,\ldots,2 {n}+ N}
 & [g_j]_{j=0,\ldots,2 {n}+N} \\
\left[-g_k \right]_{k=0,\ldots,2 {n}+N} & 0 \end{array} \right], 
\label{pevenB}
\eeq
where
\begin{equation}
F_{jk} = \left\{ \begin{array}{ll} 
{\rm sgn}(k - j) & 
(j,k = 0,\ldots,2 {n}) \\  
1& 
(j=0,\ldots,2 {n};\ 
k = 2 {n} + 1,\ldots,2 {n} + N)
\\
-1 &
(j = 2 {n} + 1,\ldots,2 {n} + N;\ 
k=0,\ldots,2 {n})
\\
{\rm sgn}(z_k - z_j) & 
(j,k = 2 {n} + 1,\ldots,2 {n} + N)
\end{array} \right. 
\label{Fodd}
\end{equation}
and
$g_j = g_k = 1$ ($j,k=0,\ldots, 2n+N$).
Substitution of Eq.(\ref{pevenB}) into Eq.(\ref{2.47})
yields 
\beq
p(z_0,\ldots,z_{2 {n} + N}) = \prod_{j=0}^{2 {n} + N} 
\sqrt{w(z_j)} \prod_{j>k\geq 0}^{2 {n}+N} (z_j - z_k) 
{\rm Pf} \left[ 
\begin{array}{ll}[F_{jk}]_{ j,k = 0,\ldots,2 {n}+ N}
 & [g_j]_{j=0,\ldots,2 {n}+N} \\
\left[-g_k \right]_{k=0,\ldots,2 {n}+N} & 0 \end{array} 
\right] .
\eeq
The Pfaffian in the above can be represented as a 
quaternion determinant, due to 
the following theorem 
(which is essentially {\em Theorem 2}):
\par
\bigskip
\noindent
{\em Theorem 2'}
\par
\bigskip
\noindent
For even $N$, we can rewrite $p(z_0,\ldots,z_{2 {n}+N})$ as 
\beq
p(z_0,\ldots,z_{2 {n} + N}) = 
\Bigl( \prod_{j=0}^{ {n}+N/2-1} r_{j}\Bigr) s_{2 {n}+N}\,
{\rm Tdet} 
[f_{jk}^{\rm even}(z_j,z_k)]_{j,k = 0,\ldots,2 {n}+N}.  
\eeq
The quaternion elements are represented as
\begin{equation}
f_{jk}^{\rm even}(z_j,z_k) = \left[ \begin{array}{cc} 
S^{\rm even}(z_j,z_k) 
& I^{\rm even}(z_j,z_k) \\ 
D^{\rm even}(z_j,z_k) & S^{\rm even}(z_k,z_j) \end{array} \right].
\eeq
and $s_j$ is defined in Eq.(\ref{sn}).
The functions $S^{\rm even}$, $D^{\rm even}$ and $I^{\rm even}$ are 
given in terms of $S$, $D$ and $I$ defined
in Eq.(\ref{SDI}), according to
\begin{eqnarray}
{S^{\rm even}}(z_j,z_k) & \!=\! & S(z_j,z_k) \Big|_\# 
\!\!+ {{\Psi}_{2 {n} + N}(z_k) \over s_{2 {n} + N}},
\nonumber\\
{D^{\rm even}}(z_j,z_k) & \!= \!& D(z_j,z_k) \Big|_\#,
\label{SDIeven}
\\
{I^{\rm even}}(z_j,z_k)  & \!=\! & I(z_j,z_k) \Big|_\# 
\!\!+ 
{{\Phi}_{2 {n} + N}(z_j) - {\Phi}_{2 {n} + N}(z_k) 
\over s_{2n+N}} .  
\nonumber
\end{eqnarray}
Here $\#$ stands for a set of substitutions
\beq
{R}_j(z) \mapsto 
R_j(z)-\frac{s_j}{s_{2n+N}}R_{2n+N}(z)
\ \ \ \ \ (j=0,\ldots,2 {n} + N-1).
\eeq
\par
\bigskip
It is straightforward to show that the quaternion element 
$f_{jk}^{\rm even}(z_j,z_k)$ in {\em Theorem 2'}
satisfies the conditions imposed 
on $f_{jk}(z_j,z_k)$ in {\em Theorem 3}. 
This means that we can inductively write
\beq
\Xi_p(z_0,\ldots,z_{2n+p}) =
\frac{(\prod_{j=0}^{p+N/2-1} r_{j}) s_{2n+N} }{
\prod_{j=0}^{2n} 
\sqrt{w(z_j)} 
\prod_{j > k\geq 0}^{2n} ( z_j - z_k )}
  {\rm Tdet}[{f_{jk}^{\rm even}}(z_j,z_k)]_{j,k=0,\ldots,2n+p}.
\eeq
Then the $p$-level correlation function 
(\ref{rho_pb})
is written as
\beq
\rho(x_1,\ldots,x_p;\{m\})  =
\frac{\Xi_p(z_0,\ldots,z_{2n+p})}{\Xi_0(z_0,\ldots,z_{2n})}
=\frac{{\rm Tdet}[f_{jk}^{\rm even}
(z_j,z_k)]_{j,k=0,\ldots,2 {n}+p}}{{\rm 
Tdet}[f_{jk}^{\rm even}(z_j,z_k)]_{j,k=0,\ldots,2 {n}}}.
\eeq

We introduce a set of notations 
\begin{eqnarray}
&&
S^{II}_{jk} = S^{\rm even}(z_j,z_k) 
\ \ \ \ \ \ \ \ \ \ \ \ 
(j,k = 0,\ldots, 2 {n}), 
\nonumber\\ 
&& 
S^{IR}_{jk} = S^{\rm even}(z_j,z_{2 {n} + k}) 
\ \ \ \ \ \ \ (j = 0,\ldots, 2 {n} ;\ k = 1,\ldots,p),
\nonumber\\
&&
S^{RI}_{jk} = S^{\rm even}(z_{2 {n} + j},z_k) 
\ \ \ \ \ \ \ 
(j = 1,\ldots,p ;\  k= 0,\ldots, 2 {n}) ,
\label{2.58}\\  
&&
S^{RR}_{jk} = S^{\rm even}(z_{2 {n} + j},z_{2 {n} + k}) 
\ \ (j,k = 1,\ldots, p)
\nonumber
\end{eqnarray}
and similarly for $D$ and $I$.
A simplification
\bea
&&S^{II}_{jk} = S^{\rm even}(-\infty,z_k) =
\frac{\Psi_{2n+N}(z_k)}{s_{2n+N}}\equiv
 S^{I}_k,  \nonumber\\ 
 &&S^{IR}_{jk} = S^{\rm even}(-\infty,z_{2n+k}) =
\frac{\Psi_{2n+N}(z_{2n+k})}{s_{2n+N}}\equiv
 S^{R}_k,
\\
&&I^{IR}_{jk} = - I^{RI}_{kj} = I^{\rm even}(-\infty,z_{2n+k}) = 
-\frac{\Phi_{2n+N}(z_{2n+k})}{s_{2n+N}}\equiv
I^{R}_k, 
 \nonumber
\eea
results from the definitions
(\ref{SDI}) and (\ref{SDIeven}).
Using Dyson's identity (\ref{Dysoneq})
we can rewrite the correlation functions as 
\begin{eqnarray}
\rho(x_1,\ldots,x_p;\{m\})  
&=&(-1)^{p(p-1) / 2} \frac{
{\rm Pf} \left[ \begin{array}{cccc}
-I^{II} & S^{II} & -I^{IR} & S^{IR} \\ 
-(S^{II})^{T} & D^{II} & -(S^{RI})^{T} & D^{IR} \\  
-I^{RI} & S^{RI} & -I^{RR} & S^{RR} \\ 
-(S^{IR})^{T} & D^{RI} & -(S^{RR})^{T} & D^{RR} \end{array} \right]  
}{{\rm Pf} \left[ \begin{array}{cc}
-I^{II} & S^{II} \\ 
-(S^{II})^{T} & D^{II} 
\end{array} \right]} \nonumber \\ 
&  =& (-1)^{p(p-1) / 2} \frac{
{\rm Pf} \left[ \begin{array}{cccc} 
 0 & S^{I} & - I^{R} & S^{R} \\ 
 - (S^{I})^{T} & D^{II} & -(S^{RI})^{T} & -(D^{RI})^T \\  
 (I^{R})^{T} & S^{RI} & -I^{RR} & S^{RR} \\ 
 - (S^{R})^{T} & D^{RI} & -(S^{RR})^{T} & 
D^{RR} \end{array} \right]  
}{{\rm Pf} \left[ \begin{array}{cc} 
0 & S^{I} \\ - (S^{I})^{T} &  D^{II} \end{array} \right]}. 
\label{2.63new}
\end{eqnarray}
In the last line we have exploited a Pfaffian identity 
that holds for antisymmetric matrices $A$, $B$ of odd ranks and
a row vector $v$:
\beq
{\rm Pf}
\left[
\ba{c|c}
A &  
\ba{c}
v\\
\vdots\\
v
\ea
\\
\hline
-v^T \cdots -v^T & B
\ea
\right]
=
{\rm Pf}
\left[
\ba{c|c}
A &  
\ba{c}
1\\
\vdots\\
1
\ea
\\
\hline
-1 \cdots -1 & 0
\ea
\right]
{\rm Pf}
\left[
\ba{cc}
0 & v\\
-v^T & B
\ea
\right]  .
\eeq

Likewise previous Subsection, 
we proceed to evaluate the component functions
of the quaternion kernel $f^{\rm even}_{jk}(z_j,z_k)$ in
the asymptitic limit where
unfolded microscopic variables 
\begin{eqnarray}
\sqrt{2N}z_{0}&\equiv& \z_0=0,\nonumber\\
\sqrt{2N}z_{2 j - 1}&\equiv&
\z_{2 j - 1}  =  i \m_j \ \ \ \ (j=1,\ldots,{n}), \nonumber \\ 
\sqrt{2N}z_{2 j }&\equiv&
\z_{2 j }  =  -i \m_j \ \ \ \ \ (j=1,\ldots,{n}), 
\label{2.63}\\ 
\sqrt{2N}z_{2n+ j }&\equiv&
\l_j 
\ \ \ \ \ \ \ \ \ \ \ \ \ \ \ \ (j= 1,\ldots,p),\nonumber
\end{eqnarray}
are kept fixed.
We note that all elements of
the sub-matrices that appear in the second line of Eq.(\ref{2.63new})
are expressed as (derivatives or integrals of)
analytic functions $\bar{S}(z,z')$ and $\bar{\Phi}_j(z)$ 
defined in Eqs.(\ref{Szz}) and (\ref{barPhi}), and $\Psi_j(z)$:
\bml
\bea
D^{II}_{jk} &=& \frac{1}{2} \frac{\partial}{\partial z_j} 
{\bar S}(z_j,z_k) ,\\
S^{RI}_{jk} & = & {\bar S}(x_j,z_k)  
 +  \frac{{\bar \Phi}_{2 \alpha + N}(x_j)}{s_{2  \alpha + N}} 
{\bar S}(-\infty,z_k)  
-  \frac{\Psi_{2 \alpha + N}(z_k) }{s_{2 \alpha + N}} 
\left( 2 \int_{-\infty}^{x_j} {\bar S}(-\infty,x) {d}x 
- 1 \right), \\  
D^{RI}_{jk} &=& 
 \frac{1}{2} 
\frac{\partial}{\partial x_j} 
{\bar S}(x_j,z_k), \\
S^{RR}_{jk} & = & {\bar S}(x_j,x_k)  
 +  \frac{{\bar \Phi}_{2 \alpha + N}(x_j) }{s_{2  \alpha + N}} 
{\bar S}(-\infty,x_k) 
- \frac{\Psi_{2 \alpha + N}(x_k) }{s_{2 \alpha + N}} 
\left( 2 \int_{-\infty}^{x_j} {\bar S}(-\infty,x) {d}x 
- 1 \right), \\ 
D^{RR}_{jk}& =& \frac{1}{2} 
\frac{\partial}{\partial x_j} 
{\bar S}(x_j,x_k) ,\\
I^{RR}_{jk} & = & - 2 \int_{x_j}^{x_k} 
{\bar S}(x_j,x) {d}x 
- {\rm sgn}(x_j - x_k) 
\nonumber\\
&&
+ \frac{\bar{\Phi}_{2 \alpha + N}(x_k) }{s_{2 \alpha + N}} 
\left( 2 \int_{-\infty}^{x_j} {\bar S}(-\infty,x) {d}x 
- 1 \right)
- \frac{\bar{\Phi}_{2 \alpha + N}(x_j) }{s_{2 \alpha + N}} 
\left( 2 \int_{-\infty}^{x_k} {\bar S}(-\infty,x) {d}x 
- 1 \right) .
\eea
\eml
Utilizing an identity 
\begin{equation}
\int_0^{\infty} {\rm e}^{-u^2/2} H_{2 k}(u) {d}u  
= 2^{2 k-1/2} \Gamma(k  + \frac12) , 
\end{equation}
an asymptotic formula
\begin{eqnarray}
\int_0^{\infty} {\rm e}^{- u^2/2} H_{2 k - 1}(u) {d}u
& = & \frac{2^{2 k - 1} \Gamma(k)}{\sqrt{\pi}} 
\sum_{\ell=0}^{k-1} 
(-1)^\ell \frac{\Gamma(\ell + 1/2)}{\Gamma(\ell + 1)} \nonumber \\ 
& = & \frac{2^{2 k - 1} \Gamma(k)}{\sqrt{\pi}}  
\left( \sqrt{\pi}\, {}_2F_1(1,\frac12;1;-1) - 
 \sum_{\ell=k}^{\infty} 
(-1)^\ell \frac{\Gamma(\ell + 1/2)}{\Gamma(\ell + 1)} \right)
\nonumber \\ 
& \sim & \frac{2^{2 k - 1} \Gamma(k)}{\sqrt{\pi}} 
\left( \sqrt{\frac{\pi}{2}} -  \frac{(-1)^k}{2 \sqrt{k}} \right)
\ \ \ \ \ \ \ (k\gg 1),
\end{eqnarray} 
and Eq.(\ref{Hermiteasy}),
we obtain Eqs.(\ref{SDI1}) 
(with $S$, $D$, $I$ in the second places replaced by 
$S^{\rm even}$, $D^{\rm even}$, $I^{\rm even}$)
and
\bml
\bea
&&{\bf S}^I_j \equiv S^I(z_j)
\sim \frac{(-1)^{n+N/2} }{\sqrt{2\pi}}\cos \z_j ,
\\
&&{\bf S}^R_j \equiv S^R(x_{j})
\sim \frac{(-1)^{n+N/2} }{\sqrt{2\pi}}\cos \l_{j},
\\
&&{\bf I}^R_j \equiv \sqrt{2N} I^R(x_{j})
\sim (-1)^{n+N/2+1}  \sqrt{\frac{2}{\pi}}\sin \l_{j}.
\eea
\label{SSI}
\eml
\noindent
These matrix elements constitute
the finite-volume partition function
\begin{eqnarray}
{\cal Z}(\{\m\})
&\equiv& 
\Xi_0(\{ \frac{\m}{\sqrt{2N}} \} )
\nonumber\\
&=&{\rm const.}\frac{{\rm Pf} \left[ \begin{array}{cc} 
0 & {\bf S}^{I} \\ - ({\bf S}^{I})^{T} &  {\bf D}^{II}
\end{array} \right]
}{
\prod_{j=1}^n \m_j^3 \prod_{j>k}^n (\m_j^2-\m_k^2)^2},
\label{calZ1b}
\eea
and the scaled spectral correlation functions
\begin{eqnarray}
\rho_s(\l_1,\ldots,\l_p;\{\m\})
&\equiv &
\Bigl(\frac{1}{\sqrt{2N}}\Bigr)^p
\rho(\frac{\l_1}{\sqrt{2N}},\ldots,\frac{\l_p}{\sqrt{2N}};
\{ \frac{\m}{\sqrt{2N}}\})
\nonumber \\ 
&=& (-1)^{p(p-1) / 2} \frac{
{\rm Pf} \left[ \begin{array}{cccc} 
 0 & {\bf S}^{I} & - {\bf I}^{R} & {\bf S}^{R} \\ 
 - ({\bf S}^{I})^{T} & {\bf D}^{II} & -({\bf S}^{RI})^{T} 
 & -({\bf D}^{RI})^T \\  
 ({\bf I}^{R})^{T} & {\bf S}^{RI} & -{\bf I}^{RR} & {\bf S}^{RR} \\ 
 - ({\bf S}^{R})^{T} & {\bf D}^{RI} & -({\bf S}^{RR})^{T} & 
{\bf D}^{RR} 
\end{array} \right]  
}{{\rm Pf} \left[ \begin{array}{cc} 
0 & {\bf S}^{I} \\ - ({\bf S}^{I})^{T} &  
{\bf D}^{II} \end{array} \right]} .
\nonumber\\
&&
\end{eqnarray}
It satisfies a sequence
\beq
\rho_s(\{\l\};\mu_1,\ldots,\mu_n,-\m_1,\ldots,-\mu_n,0)
\stackrel{\mu_{n} \to \infty}{\longrightarrow}
\rho_s(\{\l\};\mu_1,\ldots,\mu_{n-1},-\m_1,\ldots,-\mu_{n-1},0)
\stackrel{\mu_{n-1} \to \infty}{\longrightarrow}
\cdots ,
\eeq
as each of the masses is decoupled by going to infinity.
To illustrate this decoupling, we exhibit in FIG.2 a plot of
the spectral density $\rho_s(\l;\m,-\m,0)$ 
($p=1, n=1$).
\subsection{odd ${\bf N}_{\bf f}$, odd ${\bf N}$}
We finally consider the case with $N_f\equiv 2n+1$
flavors,
$\{m\}=(m_1,\ldots,m_n,-m_1,\ldots,-m_n,0)$, and with odd $N$.
As mentioned in Introduction, this case is pathological because
\beq
\Xi_p(-x_1,\ldots,-x_p;\{m\})=-\Xi_p(x_1,\ldots,x_p;\{m\})
\eeq
and namely $\Xi_0(\{m\})=Z(\{m\})=0$.
The quantities computed below
should not be considered as 
correlation functions of a RME,
but merely as multiple integrals defined by
Eqs.(\ref{2.45})$\sim$(\ref{Xi_pb}).

As in previous Subsection,
we define variables $z_j$ by Eq.(\ref{2.46}),
and denote the integrand in $\Xi_p$ as
\beq
p(z_0,\ldots,z_{2 {n} + N}) =
\prod_{j=0}^{2 {n} + N} \sqrt{w(z_j)} 
\prod_{j > k\geq 0}^{2 {n} + N} (z_j - z_k) 
\prod_{j > k > 2 {n}}^{2 {n} + N} {\rm sgn}(z_j - z_k).
\label{2.67}
\eeq
An identity
\begin{equation}
\prod_{j>k\geq 0}^{2 {n} + N} {\rm sgn}(z_j - z_k) = 
{\rm Pf}[{\rm sgn}(z_k - z_j)]_{j,k = 0,\ldots,2 {n} + N}
\end{equation}
holds for {\em real} $z_0, z_1, \ldots, z_{2 {n} + N}$. 
By taking the limit 
$z_0 < z_1 < \ldots < z_{2 {n}} \rightarrow - \infty$, 
we find another identity
\begin{equation}
\prod_{j>k>2 {n}}^{2 {n} + N} {\rm sgn}(z_j - z_k) = 
{\rm Pf}[F_{jk}]_{j,k = 0,\ldots,2 {n} + N},
\label{2.69}
\end{equation}
where
$F_{jk}$ is defined in Eq.(\ref{Fodd}).
Substitution of Eq.(\ref{2.69}) into Eq.(\ref{2.67}) yields 
\beq
p(z_0,\ldots,z_{2 {n} + N}) = 
 \prod_{j=0}^{2 {n}+N} 
\sqrt{w(z_j)} \prod_{j>k\geq 0}^{2 {n}+N} (z_j - z_k) 
{\rm Pf}[F_{jk}]_{j,k = 0,\ldots,2 {n}+N}.
\eeq
The Pfaffian in the above can be represented as
a quaternion determinant, due to the following
theorem (which is essentially {\em Theorem 1})
\par
\bigskip
\noindent
{\em Theorem 1'}
\par
\bigskip
\noindent
For odd $N$, we can rewrite $p(z_0,\ldots,z_{2 {n}+N})$ as 
\beq
p(z_0,\ldots,z_{2{n} + N}) =
\Bigl(\prod_{j=0}^{{n}+(N+1)/2-1}\!\!   r_{j}\Bigr)
{\rm Tdet} [f_{jk}(z_j,z_k)]_{j,k = 0,\ldots,2 {n}+N}.
\eeq
The quaternion elements $f_{jk}(z_j,z_k)$ are represented as
\begin{equation}
f_{jk}(z_j,z_k)= \left[ \begin{array}{cc}
S(z_j,z_k) & I(z_j,z_k) \\ 
D(z_j,z_k) & S(z_k,z_j) \end{array} \right].
\eeq
The functions $S(z_j,z_k)$, $D(z_j,z_k)$ and $I(z_j,z_k)$ 
are given by
\bea
S(z_j,z_k) &=& \sum_{\ell=0}^{{n}+(N+1)/2-1}
\frac{
\Phi_{2\ell}(z_j) \Psi_{2\ell+1}(z_k) -
\Phi_{2\ell+1}(z_j) \Psi_{2\ell}(z_k)}{r_{\ell}},
\nonumber\\
D(z_j,z_k) &=& \sum_{\ell=0}^{{n}+(N+1)/2-1}
\frac{
\Psi_{2\ell}(z_j) \Psi_{2\ell+1}(z_k) -
\Psi_{2\ell+1}(z_j) \Psi_{2\ell}(z_k)}{r_{\ell}},
\label{SDIc}\\
I(z_j,z_k) &=& -\sum_{\ell=0}^{{n}+(N+1)/2-1}
\frac{
\Phi_{2\ell}(z_j) \Phi_{2\ell+1}(z_k) - 
\Phi_{2\ell+1}(z_j) \Phi_{2\ell}(z_k)}{r_{\ell}} 
+F_{jk}.
\nonumber
\eea
\par
\bigskip
As before, the above quaternion elements
satisfy the conditions imposed 
on $f_{jk}(z_j,z_k)$ in {\em Theorem 3},
so that we can inductively write
\beq
\Xi_p(z_0,\ldots,z_{2n+p}) =
\frac{\prod_{j=0}^{p+(N+1)/2-1} r_{j} }{\prod_{j=0}^{2n} 
\sqrt{w(z_j)} 
\prod_{j > k\geq 0}^{2n} ( z_j - z_k )}
{\rm Tdet}[f_{jk}(z_j,z_k)]_{j,k=0,\ldots,2n+p} .
\eeq

In order to circumvent the vanishing of the partition function, 
we regard $z_0,z_1,\ldots,z_{2n}$ as generic 
variables, and define the 
`$p$-level correlation function'
\beq
\rho(x_1,\cdots,x_{p};\{m\})  \equiv 
 \frac{\Xi_p(z_0,\cdots,z_{2n+p})}{\Xi_0(z_0,\cdots,z_{2n})} 
=  \frac{{\rm Tdet}[f_{jk}(z_j,z_k)]_{j,k=0,\cdots,2 n+p}}{{\rm 
Tdet}[f_{jk}(z_j,z_k)]_{j,k=0,\cdots,2 n}}.
\eeq

We introduce a set of notations 
\begin{eqnarray}
&&
S^{II}_{jk} = S(z_j,z_k) 
\ \ \ \ \ \ \ \ \ \ \ \ 
(j,k = 0,\ldots, 2 {n}), 
\nonumber\\ 
&& 
S^{IR}_{jk} = S(z_j,z_{2 {n} + k}) 
\ \ \ \ \ \ \ 
(j = 0,\ldots, 2 {n} ;\ k = 1,\ldots,p),
\nonumber\\
&&
S^{RI}_{jk} = S(z_{2 {n} + j},z_k) 
\ \ \ \ \ \ \ 
(j = 1,\ldots,p ;\ k= 0,\ldots, 2 {n}) ,
\nonumber\\  
&&
S^{RR}_{jk} = S(z_{2 {n} + j},z_{2 {n} + k}) 
\ \ (j,k = 1,\ldots, p)
\label{2.58c}
\end{eqnarray}
and similarly for $D$ and $I$.
A simplification
\bea
&&S^{II}_{jk} = S^{II}(-\infty,z_k)\equiv
 S^{I}_k,  \nonumber\\ 
&&I^{IR}_{jk} = - I^{RI}_{kj} = I^{IR}(-\infty,z_{2n+k})\equiv
I^{R}_k, 
\\
&&S^{IR}_{jk} = S^{IR}(-\infty,z_{2n+k}) \equiv
 S^{R}_k, \nonumber
\eea
results again from the definition (\ref{SDIc}).
Using Dyson's identity (\ref{Dysoneq})
we can rewrite the `correlation functions' as 
Eq.(\ref{2.63new}).

Likewise two previous Subsections, 
we proceed to evaluate the component functions
of the quaternion kernel $f_{jk}(z_j,z_k)$ in
the asymptitic limit where unfolded microscopic variables
(\ref{2.63}) are kept fixed.
We note that all elements (other than 
$S^I$, $S^R$, and $I^R$) of
the sub-matrices that appear in the second line of Eq.(\ref{2.63new})
are expressed as (derivatives or integrals of)
an analytic function $\bar{S}(z,z')$ defined in Eq.(\ref{Szz})
(with $N$ replaced by $N+1$), with real and/or imaginary arguments:
\bml
\bea
D^{II}_{jk} &=& \frac{1}{2} \frac{\partial}{\partial z_j} 
{\bar S}(z_j,z_k) ,\\
S^{RI}_{jk} &=& {\bar S}(x_j,z_k) ,\\
D^{RI}_{jk} &=& 
\frac{1}{2} 
\frac{\partial}{\partial x_j} 
{\bar S}(x_j,z_k) ,\\
S^{RR}_{jk} &= &{\bar S}(x_j,x_k) ,\\
D^{RR}_{jk}& =& \frac{1}{2} 
\frac{\partial}{\partial x_j} 
{\bar S}(x_j,x_k) ,\\
I^{RR}_{jk} &=& - 2 \int_{x_j}^{x_k} 
{\bar S}(x_j,x) {d}x 
- {\rm sgn}(x_j - x_k) .
\eea
\eml
\noindent
We obtain Eqs.(\ref{SDI1}) and
\bml
\bea
&&{\bf S}^I_j \equiv S^I(z_j)
\sim \frac{(-1)^{n+(N+1)/2} }{\sqrt{2\pi}}\sin \z_j ,
\\
&&{\bf S}^R_j \equiv S^R(x_{j})
\sim \frac{(-1)^{n+(N+1)/2} }{\sqrt{2\pi}}\sin \l_{j},
\\
&&{\bf I}^R_j \equiv \sqrt{2N} I^R(x_{j})
\sim (-1)^{n+(N+1)/2}  \sqrt{\frac{2}{\pi}}\cos \l_{j}.
\eea
\label{SSIc}
\eml
\noindent
Note the phase shifts of the trigonometric functions
between Eq.(\ref{SSIc}) and its even-$N$ counterpart (\ref{SSI}).
These matrix elements constitute
the `finite-volume partition function'
\begin{eqnarray}
{\cal Z}(\{\m\})
&\equiv& 
\Xi_0(\{ \frac{\m}{\sqrt{2N}} \} )
\nonumber\\
&=&{\rm const.}\frac{{\rm Pf} \left[ \begin{array}{cc} 
0 & {\bf S}^{I} \\ - ({\bf S}^{I})^{T} &  {\bf D}^{II}
\end{array} \right]
}{
\prod_{j=1}^n \m_j^3 \prod_{j>k}^n (\m_j^2-\m_k^2)^2},
\label{calZ1c}
\eea
and
the `scaled correlation functions'
\begin{eqnarray}
\rho_s(\l_1,\ldots,\l_p;\{\m\})
&\equiv &
\Bigl(\frac{1}{\sqrt{2N}}\Bigr)^p
\rho(\frac{\l_1}{\sqrt{2N}},\ldots,\frac{\l_p}{\sqrt{2N}};
\{ \frac{\m}{\sqrt{2N}}\})
\nonumber \\ 
&=& (-1)^{p(p-1) / 2} \frac{
{\rm Pf} \left[ \begin{array}{cccc} 
 0 & {\bf S}^{I} & - {\bf I}^{R} & {\bf S}^{R} \\ 
 - ({\bf S}^{I})^{T} & {\bf D}^{II} & -({\bf S}^{RI})^{T} 
 & -({\bf D}^{RI})^T \\  
 ({\bf I}^{R})^{T} & {\bf S}^{RI} & -{\bf I}^{RR} & {\bf S}^{RR} \\ 
 - ({\bf S}^{R})^{T} & {\bf D}^{RI} & -({\bf S}^{RR})^{T} & 
{\bf D}^{RR} 
\end{array} \right]  
}{{\rm Pf} \left[ \begin{array}{cc} 
0 & {\bf S}^{I} \\ - ({\bf S}^{I})^{T} &  
{\bf D}^{II} \end{array} \right]} .
\end{eqnarray}
We again remind the reader that under the identification (\ref{2.63}),
the above `finite-volume partition function' vanishes and
the `correlation function' diverges
as its Pfaffian denominator vanishes.
\setcounter{equation}{0}
\section{symplectic ensemble}
For $\beta=4$, we concentrate on even flavor cases
for a technical reason, 
and treat the following two cases separately:
\begin{eqnarray*}
&&{\bf A}: \{m\}= 
(m_1,m_1,\ldots,m_\a,m_\a,-m_1,-m_1,\ldots,-m_\a,-m_\a),\\
&&{\bf B}: \{m\}=
(m_1,m_1,\ldots,m_\a,m_\a,-m_1,-m_1,\ldots,-m_\a,-m_\a,0,0).
\end{eqnarray*}
The $\beta=4$ case with an odd number of fermions 
will not be treated in this Article.
\subsection{${\bf N}_{\bf f}$ = 0 mod 4}
We first consider the case with $N_f\equiv 4\a$
flavors and 
$\{m\}$$=$%
$(m_1$,$m_1$,$\ldots$,$m_\a$,$m_\a$,%
$-m_1$,$-m_1$,$\ldots$,$-m_\a$,$-m_\a)$.
\linebreak[3]
We express the partition function (\ref{ZRME})
of the RME in terms of eigenvalues $\{x_j\}$ 
of $H$:
\begin{eqnarray}
&&{Z} (\{ m \})=\frac{1}{N!}
\int_{-\infty}^\infty 
\!\!\!\!\cdots \int_{-\infty}^\infty 
\prod_{j=1}^N dx_j
\prod_{j=1}^N
\biggl(
{\rm e}^{-2x_j^2}  
\prod_{k=1}^\a (x_j^2+m_k^2)
\biggr)
\prod_{j>k}^N (x_j-x_k)^4 .
\label{3.3}
\end{eqnarray}
The $p$-level correlation function of
the matrix $H$ is defined as
\bea
\rho(x_1,\ldots,x_p;\{m\})
&=&
\langle \prod_{j=1}^p {\rm tr}\,\delta(x_j-H) \rangle
\nonumber\\
&=&
\frac{\Xi_p (x_1,\ldots,x_p;\{ m \})}{\Xi_0 (\{ m \})},
\label{rho_p4}\\
\Xi_p (x_1,\ldots,x_p;\{ m \})&=&
\frac{1}{(N-p)!}
\int_{-\infty}^\infty \!\!\!\!\cdots \int_{-\infty}^\infty 
\prod_{j=p+1}^N dx_j 
\prod_{j=1}^N
\biggl(
{\rm e}^{-2x_j^2}  
\prod_{k=1}^\a (x_j^2+m_k^2)
\biggr)
\prod_{j>k}^N (x_j-x_k)^4 
\label{Xi_p4}
\eea
$(\Xi_0=Z)$.
We define new variables $z_j$ as 
\beq
\left.
\ba{l}
z_{2j-1} = i m_j\\ 
z_{2j} =  -i m_j
\ea
\right\}
\ \ \ (j=1,\ldots,\alpha). 
\eeq
Then the multiple integral (\ref{Xi_p4}) is expressed as
\bea
&&
\Xi_p(x_1,\ldots,x_p;\{z\}) = 
\frac{1}{\prod_{j=1}^{2 \alpha} w(z_j)
\prod_{j > k}^{2 \alpha} (z_j - z_k)}
\nonumber\\
&&\times
\frac{1}{(N-p)!} 
\int_{-\infty}^{\infty} \!\!\!\cdots \int_{-\infty}^{\infty} 
\prod_{j=2\a+p+1}^{2\a+N} {d}x_{j} 
\prod_{j=1}^{2 \alpha} w(z_j) 
\prod_{j=1}^N w(x_j)^2 
\prod_{j > k}^{2 \alpha} (z_j - z_k) 
\prod_{j=1}^N 
\prod_{k=1}^{2 \alpha} (x_j - z_k)^2 
\prod_{j > k}^{N} (x_j - x_k)^4 ,
\label{1A11}
\eea
where
$w(z) = {\rm e}^{- z^2}.$

Let us denote the integrand in Eq.(\ref{1A11}) as
\beq
p(z_1,\ldots,z_{2 \alpha};x_1,\ldots,x_N) = 
\prod_{j=1}^{2 \alpha} w(z_j) 
\prod_{j=1}^N w(x_j) ^2 
\prod_{j > k}^{2 \alpha} (z_j - z_k) 
\prod_{j=1}^N 
\prod_{k=1}^{2 \alpha} (x_j - z_k)^2 
\prod_{j > k}^{N} (x_j - x_k)^4 .
\eeq
The above expressions can be 
represented as a quaternion determinant.
In doing so, we introduce 
monic skew-orthogonal polynomials 
$Q_j(z)=z^j+\cdots$ that satisfy
\bea
&&\langle Q_{2j}, Q_{2k+1} \rangle_Q =
 - \langle Q_{2k+1}, Q_{2j}\rangle_Q = q_j \delta_{jk},
\label{skew4}\\
&&\langle Q_{2j}, Q_{2k} \rangle_Q =
 \langle Q_{2j+1}, Q_{2k+1}\rangle_Q = 0,
 \nonumber
\eea
where
\begin{equation} 
\langle f, g \rangle_Q = 
\int_{-\infty}^{\infty} dz\,w(z)^2 \bigl( 
f(z) g'(z) - f'(z) g(z) \bigr) .
\end{equation}
Explicit forms for the skew-orthogonal polynomials and their norms
associated with the Gaussian weight $w(z)$
are known in terms of the Hermite 
polynomials
\cite{Meh}:
\begin{eqnarray}
Q_{2 j}(z) & = & - \frac{1}{ 2^{3j+1/2}} {\rm e}^{z^2} 
\int_{-\infty}^z {\rm e}^{-z'{}^{2}} H_{2 j + 1}(\sqrt{2} z') {d}z' , 
\nonumber \\
Q_{2 j + 1}(z) & = & \frac{1}{ 2^{3 j + 3/2}} 
H_{2 j + 1}(\sqrt{2} z), \\
q_j &=& 2^{-4 j-1/2} \sqrt{\pi} (2j + 1)! ~. \nonumber
\eea
It can be readily seen that 
(cf.\ Eq.(26) of Ref.\cite{NN1})
$p(z_1,\ldots,z_{2 \alpha};x_1,\ldots,x_N)$ is represented
as a Pfaffian:
\begin{eqnarray}
& & p(z_1,\ldots,z_{2 \alpha};x_1,\ldots,x_N) 
= \det \left[ \begin{array}{cccc} 
\Psi_0(z_1) & \Psi_1(z_1) & \cdots & \Psi_{2 N + 2 \alpha - 1}(z_1) \\ 
 \vdots & \vdots & \ddots & \vdots \\  
\Psi_0(z_{2 \alpha}) & \Psi_1(z_{2 \alpha}) & \cdots & 
\Psi_{2 N + 2 \alpha - 1}(z_{2 \alpha}) \\  
\Psi_0^{\prime}(x_1) & \Psi_1^{\prime}(x_1) & \cdots & 
\Psi_{2 N + 2 \alpha - 1}^{\prime}(x_1) \\  
\Psi_0(x_1) & \Psi_1(x_1) & \cdots & 
\Psi_{2 N + 2 \alpha - 1}(x_1) \\  
 \vdots & \vdots & \ddots & \vdots \\  
\Psi_0^{\prime}(x_N) & \Psi_1^{\prime}(x_N) & \cdots & 
\Psi_{2 N + 2 \alpha - 1}^{\prime}(x_N) \\  
\Psi_0(x_N) & \Psi_1(x_N) & \cdots & 
\Psi_{2 N + 2 \alpha - 1}(x_N) \end{array} \right] \nonumber \\ 
& & =\!
\Bigl(\prod_{j=0}^{\alpha + N-1} q_{j} \Bigr)
{\rm Pf}\left[ \begin{array}{ll} 
\left[ \begin{array}{cc} 
- I(z_{2 j},z_{2 k}) & I(z_{2 j},z_{2 k - 1}) \\ 
I(z_{2 j - 1},z_{2 k}) & - I(z_{2 j - 1},z_{2 k - 1}) \end{array} 
\right]_{j,k=1,\ldots,\alpha} &  \!
\left[ \begin{array}{cc} 
S(z_{2 j},x_{k}) & I(z_{2 j},x_{k}) \\ 
-S(z_{2 j - 1},x_{k}) & - I(z_{2 j - 1},x_{k}) \end{array} 
\right]_{j=1,\ldots,\alpha ; k = 1,\ldots,N} \\    
\left[ \begin{array}{cc} 
- S(z_{2 k},x_{j}) & S(z_{2 k-1},x_{j}) \\ 
I(x_{j},z_{2 k}) & - I(x_{j},z_{2 k - 1}) \end{array} 
\right]_{j=1,\ldots,N; k=1,\ldots,\alpha} &  \! 
\left[ \begin{array}{cc} 
D(x_{j},x_{k}) & S(x_{k},x_{j}) \\ 
-S(x_{j},x_{k}) & - I(x_{j},x_{k}) \end{array} 
\right]_{j,k = 1,\ldots,N} 
\end{array} \right]. \nonumber \\  
\end{eqnarray} 
Here
\begin{equation}
\Psi_j(z) = w(z) Q_j(z),
\end{equation}
and
the functions $S(x,y)$, $D(x,y)$, $I(x,y)$ are given by 
\bea
&&S(x,y) = \sum_{j=0}^{\alpha + N-1} \frac{ 
\Psi_{2 j}(x) \Psi'_{2 j+1}(y) - 
\Psi_{2 j+1}(x) \Psi'_{2 j}(y)}{q_j} , 
\nonumber\\
&&D(x,y) = \sum_{j=0}^{\alpha + N-1} \frac{ 
\Psi'_{2 j}(x) \Psi'_{2 j+1}(y) - 
\Psi'_{2 j+1}(x) \Psi'_{2 j}(y)}{q_j}
= \frac{\partial}{\partial x}S(x,y) ,
\label{3.13}\\
&&I(x,y) =- \sum_{j=0}^{\alpha + N-1} \frac{ 
\Psi_{2 j}(x) \Psi_{2 j+1}(y) - 
\Psi_{2 j+1}(x) \Psi_{2 j}(y)}{q_j} 
=-\int_x^y S(x,z) dz .
\nonumber
\eea
Using Dyson's identity (\ref{Dysoneq}), we can convert the Pfaffian
into a quaternion determinant:
\begin{equation}
p(z_1,\ldots,z_{2 \alpha};x_1,\ldots,x_N) 
= \Bigl( \prod_{j=0}^{\alpha + N-1} q_{j}  \Bigr)
{\rm Tdet} \left[ 
\begin{array}{ll}
\left[g(z_{2 j - 1},z_{2 j};z_{2 k -1},z_{2 k}) 
\right]_{j,k=1,\ldots,\alpha} &   \!
\left[h(z_{2 j - 1},z_{2 j};x_k)\right]_{
j=1,\ldots,\alpha ; k = 1,\ldots,N} \\ 
\bigl[{\hat h}(z_{2 k - 1},z_{2 k};x_j)\bigl]_{
j=1,\ldots,N ; k=1,\ldots,\alpha} & \!
\left[f(x_j,x_k)\right]_{j,k = 1,\ldots,N} 
\end{array} \right],
\end{equation}
where
\begin{eqnarray}
&&g(z_{2 j - 1},z_{2 j};z_{2 k -1},z_{2 k})  = 
\left[ \begin{array}{ll} 
- I(z_{2 j - 1},z_{2 k}) & I(z_{2 j - 1},z_{2 k - 1}) \\  
- I(z_{2 j},z_{2 k}) & I(z_{2 j},z_{2 k - 1})  
\end{array} 
\right], \nonumber\\  
&&h(z_{2 j - 1},z_{2 j};x_k)  = 
\left[ \begin{array}{ll} 
S(z_{2 j - 1},x_{k}) & I(z_{2 j - 1},x_{k}) \\  
S(z_{2 j},x_{k}) & I(z_{2 j},x_{k})  
\end{array} 
\right], \\ 
&&f(x_j,x_k)  = 
\left[ \begin{array}{ll} 
S(x_{j},x_{k}) & I(x_{j},x_{k}) \\ 
D(x_{j},x_{k}) & S(x_{k},x_{j}) 
\end{array} \right], \nonumber  
\end{eqnarray} 
and $\hat{h}$ stands for the dual of $h$
(see Appendix A).
The skew-orthogonality relations (\ref{skew4}) lead to 
\bea
&&\int_{-\infty}^{\infty} 
f(x^{\prime},x) f(x,x^{\prime \prime}) {d}x 
= f(x^{\prime},x^{\prime \prime}),
\nonumber\\
&&
\int_{-\infty}^{\infty} h(z,z^{\prime};x) f(x,x^{\prime}) {d}x 
= h(z,z^{\prime};x^{\prime}),
\\
&&
\int_{-\infty}^{\infty} h(z,z^{\prime},x) 
{\hat h}(w,w^{\prime};x) {d}x 
= g(z,z^{\prime};w,w^{\prime}),
\nonumber
\eea
which mean that the condition of {\em Theorem 3} is satisfied. 
Therefore we can inductively write
\bea
&&\Xi_p(x_1,\ldots,x_p;z_1,\ldots,z_{2\a}) = 
\nonumber\\
&&\frac{
\prod_{j=0}^{\alpha + N-1} q_{j} }{\prod_{j=1}^{2 \alpha} w(z_j)
\prod_{j > k}^{2 \alpha} (z_j - z_k)}
 {\rm Tdet}
 \left[ \begin{array}{ll} [g(z_{2 j - 1},z_{2 j}
;z_{2 k -1},z_{2 k}) ]_{j,k=1,\ldots,\alpha} &   
[h(z_{2 j - 1},z_{2 j};x_k)]_{j=1,\ldots,\alpha ; k = 1,\ldots,p}
\\ 
{}[{\hat h}(z_{2 k - 1},z_{2 k};x_j)]_{
j=1,\ldots,p ; k=1,\ldots,\alpha} 
& [f(x_j,x_k)]_{j,k = 1,\ldots,p} 
\end{array} \right]   .
\eea
Then the $p$-level correlation function (\ref{rho_p4})
is written as
\bea
\rho(x_1,\ldots,x_p;\{m\})&=&
\frac{\Xi_p(x_1,\ldots,x_p;z_1,\ldots,z_{2n})
}{\Xi_0(z_1,\ldots,z_{2n})}
\nonumber\\
&=&\frac{{\rm Tdet}
 \left[ 
 \begin{array}{ll} 
[g(z_{2 j - 1},z_{2 j};z_{2 k -1},z_{2 k})]_{j,k=1,\ldots,\alpha} &   
[h(z_{2 j - 1},z_{2 j};x_k)]_{j=1,\ldots,\alpha ; k = 1,\ldots,p} \\ 
{}[{\hat h}(z_{2 k - 1},z_{2 k};x_j)]_{
j=1,\ldots,p ; k=1,\ldots,\alpha} 
& [f(x_j,x_k)]_{j,k = 1,\ldots,p} 
\end{array} \right]
}{{\rm Tdet}[g(z_{2 j - 1},z_{2 j};z_{2 k -1},z_{2 k}) ]_{
j,k=1,\ldots,\alpha}}.
\eea
We introduce a set of notations 
\begin{eqnarray}
&&S^{II}_{jk}  =  S(z_j,z_k)
\ \ \ \ \ \ (j,k = 1,\ldots, 2 \alpha),
\nonumber \\  
&&S^{IR}_{jk}  =  
S(z_j,x_k)
\ \ \ \ \ 
(
j = 1,\ldots,2 \alpha;\ 
k = 1,\ldots,p
), \\   
&&S^{RR}_{jk}  =  
S(x_j,x_k) 
\ \ \ \ (j,k = 1,\ldots,p), \nonumber
\end{eqnarray} 
and similarly for $D$ and $I$.
Using Dyson's identity (\ref{Dysoneq}) again, we can convert 
the quaternion determinant back to a Pfaffian, so that
the correlation functions read
\beq
\rho(x_1,\ldots,x_p;\{m\}) 
= 
(-1)^{p(p-1) / 2} 
\frac{
{\rm Pf} \left[ \begin{array}{ccc}
-I^{II} & -I^{IR} & S^{IR} \\ 
(I^{IR})^T &  -I^{RR} & S^{RR} \\ 
-(S^{IR})^{T} & -(S^{RR})^{T} & D^{RR} \end{array} \right]  
}{{\rm Pf}[-I^{II}]}. 
\label{Pf4}
\eeq
Now we proceed to evaluate 
the component functions
of the quaternion kernel, which are expressed as 
(derivatives or integrals of) the analytic function $S(x,y)$
with real and/or imaginary arguments,
in the asymptotic limit
where
unfolded microscopic variables 
\begin{eqnarray}
\sqrt{2N}z_{2 j - 1}&\equiv&
\z_{2j-1}= i \m_j \ \ \ (j=1,\ldots,{\a}), \nonumber \\ 
\sqrt{2N}z_{2 j }&\equiv&
\z_{2j}= -i \m_j \ \ \ \, (j=1,\ldots,{\a}), \\ 
\sqrt{2N}x_{j}&\equiv&
\l_j \ \ \ \ \ \ \ \ \ \ \ \ \ \ \ (j= 1,\ldots,p),\nonumber
\end{eqnarray}
are kept fixed.
We note that
$S(x,y)$ has a compact expression
\begin{eqnarray}
S(x,y) & = & 
\frac{{\rm e}^{-x^2-y^2}}{2^{2 \alpha + 2 N + 1} \sqrt{\pi} 
\Gamma(2 \alpha + 2 N)} 
\frac{H_{2 \alpha + 2 N}(\sqrt{2} x) 
H_{2 \alpha + 2 N - 1}(\sqrt{2} y) - H_{2 \alpha + 2 N - 1}(\sqrt{2} x) 
H_{2 \alpha + 2 N}(\sqrt{2} y)}{x - y} \nonumber \\ 
& + & 
\frac{{\rm e}^{-y^2}}{2^{2 \alpha + 2 N} \sqrt{\pi} 
\Gamma(2 \alpha + 2 N)} 
H_{2 \alpha + 2 N}(\sqrt{2} y) 
\int_{-\infty}^x {\rm e}^{-u^2} 
H_{2 \alpha + 2 N - 1}(\sqrt{2} u) 
{d}u.
\label{Sbar}
\end{eqnarray}
In the second line of Eq.(\ref{Sbar}),
we have singled out the unitary scalar kernel and
applied to it the Christoffel-Darboux formula.
Substituting asymptotic formulas for
the Hermite polynomials (\ref{Hermiteasy}),
one can show that ${S}(x,y)$ approaches
the sine kernel \cite{Meh}:
\beq
\frac{1}{\sqrt{2N}} 
{S}(\frac{\z}{\sqrt{2N}},\frac{\z'}{\sqrt{2N}})
\sim
\frac{\sin 2(\z-\z')}{2\pi(\z-\z')}= 
K(2(\zeta-\zeta')) .
\eeq
The Pfaffian elements in the the asymptotic limit,
after taking into account an unfolding by the factor
$\sqrt{2N}$, are then expressed in terms of $K(\z)$:
\bml
\bea
{\bf I}^{II}_{jk} 
&\equiv&
I(z_j,z_k)
\sim  \int_{0}^{\z_j-\z_k} 
\!\!\!\!\!\!\!\!\!\!\!
K(2\z){d}\z,\\
{\bf S}^{IR}_{jk} 
&\equiv&
 \frac{1}{\sqrt{2 N}} S(z_j,x_k) 
\sim K(2(\z_j-\l_k)), \\
{\bf I}^{IR}_{jk} &\equiv&
I(z_j,x_k) 
\sim \int_0^{\z_j-\l_k} \!\!\!\!\!\!\!\!\!\!\!
K(2\z){d}\z,\\
{\bf S}^{RR}_{jk} 
&\equiv& 
\frac{1}{\sqrt{2 N}} S(x_j,x_k) 
\sim K(2(\l_j-\l_k)), \\
{\bf D}^{RR}_{jk} &\equiv& 
 \frac{1}{2 N}  D(x_j,x_k) 
\sim 
2K'(2(\l_j-\l_k)),\\
{\bf I}^{RR}_{jk} 
&\equiv&
I(x_j,x_k) 
\sim
\int_{0}^{\l_j-\l_k}
\!\!\!\!\!\!\!\!\!\!\!
K(2\l){d}\l.
\eea
\label{SDI4}
\eml
\noindent
These matrix elements
constitute
the finite-volume partition function
\begin{eqnarray}
{\cal Z}(\{m\})
&\equiv &
\Xi_0(\{\frac{\m}{\sqrt{2N}}\})
\nonumber\\
&=&{\rm const.}\frac{{\rm Pf}[-{\bf I}^{II}]}{
\prod_{j=1}^\a \m_j \prod_{j>k}^\a (\m_j^2-\m_k^2)^2},
\label{calZ4}
\eea
and 
the scaled spectral correlation functions
\begin{eqnarray}
\rho_s(\l_1,\ldots,\l_p;\{\m\})
&\equiv &
\Bigl(\frac{1}{\sqrt{2N}}\Bigr)^p
\rho(\frac{\l_1}{\sqrt{2N}},\ldots,\frac{\l_p}{\sqrt{2N}};
\{
\frac{\m}{\sqrt{2N}}
\})
\nonumber \\ 
&=&  (-1)^{p(p-1) / 2} \frac{
{\rm Pf} \left[ \begin{array}{ccc}
 -{\bf I}^{II} & -{\bf I}^{IR} & {\bf S}^{IR} \\  
 ({\bf I}^{IR})^T & -{\bf I}^{RR} & {\bf S}^{RR} \\ 
 -({\bf S}^{IR})^T & -({\bf S}^{RR})^{T} & {\bf D}^{RR}
\end{array} \right]  
}{{\rm Pf} \left[ -{\bf I}^{II} \right]}.  
\label{rhos4}
\end{eqnarray}
In the quenched limit $\mu_1,\ldots,\mu_\a\to \infty$
when the ratio of two Pfaffians is replaced by a minor 
$\pf 
\left[{
\,-{\bf I}^{RR} \ \ \ \ {\bf S}^{RR}   \atop
 -({\bf S}^{RR})^{T}\ {\bf D}^{RR} }\right]
$,
the correlation functions approach
those of the Gaussian symplectic ensemble \cite{Meh}.
By the same token, it satisfies a sequence
\bea
&&
\rho_s(\{\l\};\mu_1,\mu_1,\ldots,\mu_n,\mu_n,
-\mu_1,-\mu_1,\ldots,-\mu_n,-\mu_n)
\stackrel{\mu_{n} \to \infty}{\longrightarrow}
\\
&&
\rho_s(\{\l\};\mu_1,\mu_1,\ldots,\mu_{n-1},\mu_{n-1},
-\mu_1,-\mu_1,\ldots,-\mu_{n-1},-\mu_{n-1})
\stackrel{\mu_{n-1} \to \infty}{\longrightarrow}
\cdots ,
\nonumber
\eea
as each of the masses is decoupled by going to infinity.
To illustrate this decoupling, we exhibit in FIG.3 a plot of
the spectral density $\rho_s(\l;\m,\m,-\m,-\m)$  ($p=1, \a=1$).
\subsection{${\bf N}_{\bf f}$ = 2 mod 4}
Next we consider the case with $N_f\equiv 4\a+2$
flavors and 
$\{m\}=(m_1,m_1,\ldots,m_\a,m_\a,m_1,-m_1,\ldots$, 
$-m_\a,-m_\a,0,0)$.
We express the partition function (\ref{ZRME})
of the RME in terms of eigenvalues $\{x_j\}$ 
of $H$ 
\begin{eqnarray}
&&{Z} (\{ m \})=\frac{1}{N!}
\int_{-\infty}^\infty 
\!\!\!\!\cdots \int_{-\infty}^\infty 
\prod_{j=1}^N dx_j
\prod_{j=1}^N
\biggl(
{\rm e}^{-2x_j^2} x_j^2 
\prod_{k=1}^\a (x_j^2+m_k^2)
\biggr)
\prod_{j>k}^N (x_j-x_k)^4 .
\label{3.3b}
\end{eqnarray}
The $p$-level correlation function of
the matrix $H$ is defined as
\bea
\rho(x_1,\ldots,x_p;\{m\})
&=&
\langle \prod_{j=1}^p {\rm tr}\,\delta(x_j-H) \rangle
\nonumber\\
&=&
\frac{\Xi_p (x_1,\ldots,x_p;\{ m \})}{\Xi_0 (\{ m \})},
\label{rho_p4b}\\
\Xi_p (x_1,\ldots,x_p;\{ m \})&=&
\frac{1}{(N-p)!}
\int_{-\infty}^\infty \!\!\!\!\cdots \int_{-\infty}^\infty 
\prod_{j=p+1}^N dx_j \prod_{j=1}^N
\biggl(
{\rm e}^{-2x_j^2} x_j^2 
\prod_{k=1}^\a (x_j^2+m_k^2)
\biggr)
\prod_{j>k}^N (x_j-x_k)^4
\label{Xi_p4b}
\eea
$(\Xi_0=Z)$.
We define new variables $z_j$ as 
\bea
&& z_0=0,\nonumber\\
&&
\!\!
\left.
\ba{l}
z_{2j-1} = i m_j\\ 
z_{2j} =  -i m_j
\ea
\right\}
\ \ \ (j=1,\ldots,\alpha). 
\eea
Then the multiple integral (\ref{Xi_p4b}) is expressed as
\bea
&&\Xi_p(x_1,\ldots,x_p;\{z\}) = 
\frac{1}{\prod_{j=0}^{2 \alpha} w(z_j)
\prod_{j > k\geq 0}^{2 \alpha} (z_j - z_k)}
\nonumber\\
&&\times
\frac{1}{(N-p)!} 
\int_{-\infty}^{\infty} \!\!\!\cdots \int_{-\infty}^{\infty} 
\prod_{j=2\a+p+1}^{2\a+N} {d}x_{j} 
\prod_{j=0}^{2 \alpha} w(z_j) 
\prod_{j=1}^N w(x_j)^2 
\prod_{j > k\geq 0}^{2 \alpha} (z_j - z_k) 
\prod_{j=1}^N 
\prod_{k=0}^{2 \alpha} (x_j - z_k)^2 
\prod_{j > k}^{N} (x_j - x_k)^4 ,
\label{1A11b}
\eea
where $w(z) = {\rm e}^{- z^2}$.

Let us denote the integrand in Eq.(\ref{1A11b}) as
\beq
p(z_0,\ldots,z_{2 \alpha};x_1,\ldots,x_N) = 
\prod_{j=0}^{2 \alpha} w(z_j) 
\prod_{j=1}^N w(x_j) ^2 
\prod_{j > k\geq 0}^{2 \alpha} (z_j - z_k) 
\prod_{j=1}^N 
\prod_{k=0}^{2 \alpha} (x_j - z_k)^2 
\prod_{j > k}^{N} (x_j - x_k)^4 .
\eeq

It can be readily seen that 
$p(z_0,\ldots,z_{2 \alpha};x_1,\ldots,x_N)$ is represented
as a Pfaffian:
\begin{eqnarray}
& & p(z_0,\ldots,z_{2 \alpha};x_1,\ldots,x_N) 
 =  \det \left| \begin{array}{cccc} 
\Psi_0(z_0) & \Psi_1(z_0) & \cdots & \Psi_{2 N + 2 \alpha}(z_0) \\ 
\Psi_0(z_1) & \Psi_1(z_1) & \cdots & \Psi_{2 N + 2 \alpha}(z_1) \\ 
 \vdots & \vdots & \ddots & \vdots \\  
\Psi_0(z_{2 \alpha }) & \Psi_1(z_{2 \alpha }) & \cdots & 
\Psi_{2 N + 2 \alpha}(z_{2 \alpha }) \\  
\Psi_0^{\prime}(x_1) & \Psi_1^{\prime}(x_1) & \cdots & 
\Psi_{2 N + 2 \alpha}^{\prime}(x_1) \\  
\Psi_0(x_1) & \Psi_1(x_1) & \cdots & 
\Psi_{2 N + 2 \alpha}(x_1) \\  
 \vdots & \vdots & \ddots & \vdots \\  
\Psi_0^{\prime}(x_N) & \Psi_1^{\prime}(x_N) & \cdots & 
\Psi_{2 N + 2 \alpha}^{\prime}(x_N) \\  
\Psi_0(x_N) & \Psi_1(x_N) & \cdots & 
\Psi_{2 N + 2 \alpha}(x_N) \end{array} \right| \nonumber \\ 
& = & 
\Bigl( \prod_{j=0}^{\alpha + N-1} q_{j}\Bigr)
 {\rm Pf}\left[ \begin{array}{ccc}  
[G_{jk}]_{j,k=1,\ldots,\alpha} & [A_{j}]_{j=1,\ldots,\alpha} 
& [H_{jk}]_{j=1,\ldots,\alpha; k=1,\ldots,N}  \cr 
[{\hat A}_k]_{k=1,\ldots,\alpha} & \Omega & 
[B_{k}]_{k=1,\ldots,N} \cr 
[{\hat H}_{kj}]_{j=1,\ldots,N; k=1,\ldots,\alpha} 
& [{\hat B}_{j}]_{j=1,\ldots,N} & [F_{jk}]_{j,k=1,\ldots,N}  
\end{array} \right],  
\end{eqnarray} 
where
\begin{eqnarray}
G_{jk} & = & \left[ \begin{array}{cc} 
 - {\tilde I}(z_{2 j},z_{2k}) & {\tilde I}(z_{2 j},z_{2 k - 1}) \\ 
{\tilde I}(z_{2 j - 1},z_{2k}) & 
- {\tilde I}(z_{2 j - 1},z_{2 k - 1}) \end{array} 
\right], \nonumber \\    
H_{jk} & = & \left[ \begin{array}{cc} 
{\tilde S}(z_{2 j},x_{k}) & {\tilde I}(z_{2 j},x_{k}) \\ 
-{\tilde S}(z_{2 j - 1},x_{k}) & - {\tilde I}(z_{2 j - 1},x_{k}) \end{array} 
\right], \nonumber \\      
F_{jk} & = & \left[ \begin{array}{cc} 
{\tilde D}(x_{j},x_{k}) & {\tilde S}(x_{k},x_{j}) \\ 
-{\tilde S}(x_{j},x_{k}) & - {\tilde I}(x_{j},x_{k}) \end{array} 
\right], \nonumber \\  
A_j & = & \left[ \begin{array}{cc} 
 \Psi_{2 N + 2 \alpha}(z_{2 j}) & {\tilde I}(z_{2 j},z_{0}) \\ 
- \Psi_{2 N + 2 \alpha}(z_{2 j - 1}) & 
- {\tilde I}(z_{2 j - 1},z_{0}) 
\end{array} \right], \\    
B_k & = & \left[ \begin{array}{cc} 
- \Psi_{2 N + 2 \alpha}^{\prime}(x_k) & 
\Psi_{2 N + 2 \alpha}(x_k) \\ 
-{\tilde S}(z_{0},x_{k}) & 
- {\tilde I}(z_{0},x_{k}) \end{array} 
\right], \nonumber \\  
\Omega & = & \left[ \begin{array}{cc} 
0 & \Psi_{2 N + 2 \alpha}(z_{0}) \\ 
-\Psi_{2 N + 2 \alpha}(z_{0}) & 0  \end{array} 
\right].   \nonumber
\end{eqnarray}
The functions $\tilde{S}$, $\tilde{D}$, $\tilde{I}$ are 
given in terms of $S$, $D$ and $I$ defined in Eq.(\ref{3.13})
according to
\begin{eqnarray}
\tilde{S}(x,y) & \!=\! & S(x,y) \Big|_\ast 
\!\!+ {{\Psi}_{2\a + 2N}(x) \over s_{2\a+2N}},
\nonumber\\
\tilde{D}(x,y) & \!= \!& D(x,y) \Big|_\ast ,\\
\tilde{I}(x,y)  & \!=\! & I(x,y) \Big|_\ast 
\!\!+ 
{\Psi'_{2\a + 2N}(x) - \Psi'_{2\a+2N}(y) 
\over s_{2\a+2N}} , 
\nonumber
\end{eqnarray}
and
\beq
s_j=\int_{-\infty}^\infty \Psi_j(x)dx.
\eeq
Here $\ast$ stands for a set of substitutions
\beq
{Q}_j(z) \mapsto 
Q_j(z)-\frac{s_j}{s_{2\a+2N}}Q_{2\a+2N}(z)
\ \ \ \ \ (j=0,\ldots,2\a + 2N-1).
\eeq
Using Dyson's identity (\ref{Dysoneq}), we find
\begin{eqnarray}
& & p(z_0,\ldots,z_{2 \alpha};x_1,\ldots,x_N) \nonumber \\  
& = & \Bigl( \prod_{j=0}^{\alpha + N-1} q_{j}\Bigr) 
{\rm Tdet} \left[ \begin{array}{ccc} [g(z_{2 j - 1},z_{2 j}
;z_{2k-1},z_{2k}) ]_{j,k=1,\ldots,\alpha} & 
[a(z_{2 j - 1},z_{2 j})]_{j=1,\ldots,\alpha} &   
[h(z_{2 j - 1},z_{2 j};x_k)]_{j=1,\ldots,\alpha; k = 1,\ldots,N} \cr 
[{\hat a}(z_{2 k - 1},z_{2k})]_{k=1,\ldots,\alpha} & \omega & 
[b(x_k)]_{k=1,\ldots,N} \cr 
[{\hat h}(z_{2 k - 1},z_{2k};x_j)]_{j=1,\ldots,N; k=1,\ldots,\alpha} 
& [{\hat b}(x_j)]_{j=1,\ldots,N} & [f(x_j,x_k)]_{j,k = 1,\ldots,N} 
\end{array} \right], \nonumber \\ 
\end{eqnarray} 
where
\begin{eqnarray}
g(z_{2 j - 1},z_{2 j};z_{2k-1},z_{2k}) & = &  
\left[ \begin{array}{cc} 
- {\tilde I}(z_{2 j - 1},z_{2k}) & {\tilde I}(z_{2 j - 1},z_{2 k - 1}) \\  
- {\tilde I}(z_{2 j},z_{2k}) & {\tilde I}(z_{2 j},z_{2 k - 1})  
\end{array} 
\right], \nonumber \\  
h(z_{2 j - 1},z_{2 j};x_k) & = &   
\left[ \begin{array}{cc} 
{\tilde S}(z_{2 j - 1},x_{k}) & {\tilde I}(z_{2 j - 1},x_{k}) \\  
{\tilde S}(z_{2 j},x_{k}) & {\tilde I}(z_{2 j},x_{k})  
\end{array} 
\right], \nonumber \\ 
f(x_j,x_k) & = &  
\left[ \begin{array}{cc} 
{\tilde S}(x_{j},x_{k}) & {\tilde I}(x_{j},x_{k}) \\ 
{\tilde D}(x_{j},x_{k}) & {\tilde S}(x_{k},x_{j}) 
\end{array} \right], \nonumber \\  
a(z_{2 j -1},z_{2j}) & = & \left[ \begin{array}{cc} 
 \Psi_{2 N + 2 \alpha}(z_{2 j - 1}) & 
{\tilde I}(z_{2 j - 1},z_{0}) \\  
 \Psi_{2 N + 2 \alpha}(z_{2 j}) & 
{\tilde I}(z_{2 j},z_{0})  
\end{array} \right], \\    
b(x_k) & = & \left[ \begin{array}{cc} 
 {\tilde S}(z_{0},x_{k}) & 
{\tilde I}(z_{0},x_{k}) \\  
- \Psi_{2 N + 2 \alpha}^{\prime}(x_k) & 
\Psi_{2 N + 2 \alpha}(x_k) \end{array} 
\right], \nonumber \\  
\omega & = & \left[ \begin{array}{cc} 
 \Psi_{2 N + 2 \alpha}(z_{0}) & 0 \\  
0 & \Psi_{2 N + 2 \alpha}(z_{0})  
\end{array} 
\right] ,  
\nonumber
\end{eqnarray}
and $\hat{a}$, $\hat{b}$, $\hat{h}$ stand for 
the duals of $a$, $b$, $h$.
The skew-orthogonality relations (\ref{skew4}) lead to 
\bea
&&\int_{-\infty}^{\infty} f(x^{\prime},x) f(x,x^{\prime \prime}) {d}x 
= f(x^{\prime},x^{\prime \prime}),
\nonumber\\
&&\int_{-\infty}^{\infty} b(x) f(x,x^{\prime}) {d}x 
= b(x^{\prime}),
\nonumber\\
&&
\int_{-\infty}^{\infty} h(z,z^{\prime};x) f(x,x^{\prime}) {d}x 
= h(z,z^{\prime};x^{\prime}),
\nonumber\\
&&
\int_{-\infty}^{\infty} b(x) {\hat b}(x) {d}x 
= \omega,\\
&&
\int_{-\infty}^{\infty} h(z,z^{\prime};x) {\hat b}(x) {d}x 
= a(z,z^{\prime}) ,
\nonumber\\
&&
\int_{-\infty}^{\infty} h(z,z^{\prime},x) 
{\hat h}(w,w^{\prime};x) {d}x 
= g(z,z^{\prime};w,w^{\prime}),
\nonumber
\eea
which mean that the condition of {\em Theorem 3} is satisfied. 
Therefore we can inductively write
\bea
&&
\Xi_p(x_1,\ldots,x_p;z_0,\ldots,z_{2\alpha}) = 
\frac{
\prod_{j=0}^{\alpha + N-1} q_{j} }{\prod_{j=0}^{2 \alpha} w(z_j)
\prod_{j > k\geq 0}^{2 \alpha} (z_j - z_k)}
\nonumber\\
&&
\times    
{\rm Tdet} \left[ \begin{array}{ccc} [g(z_{2 j - 1},z_{2 j}
;z_{2k-1},z_{2k}) ]_{j,k=1,\ldots,\alpha} & 
[a(z_{2 j - 1},z_{2 j})]_{j=1,\ldots,\alpha} &   
[h(z_{2 j - 1},z_{2 j};x_k)]_{j=1,\ldots,\alpha;
k = 1,\ldots,p} \cr 
[{\hat a}(z_{2 k - 1},z_{2k})]_{k=1,\ldots,\alpha} & \omega & 
[b(x_k)]_{k=1,\ldots,p} \cr 
[{\hat h}(z_{2 k - 1},z_{2k};x_j)]_{j=1,\ldots,p; k=1,\ldots,\alpha} 
& [{\hat b}(x_j)]_{j=1,\ldots,p} & [f(x_j,x_k)]_{j,k = 1,\ldots,p} 
\end{array} \right]. 
\label{Xi}
\eea
Then the $p$-level correlation function (\ref{rho_p4b})
is written as
\begin{eqnarray}
& & \rho(x_1,\ldots,x_p;\{m\}) =
\frac{\Xi_p(x_1,\ldots,x_p;z_0,\ldots,z_{2n})
}{\Xi_0(z_0,\ldots,z_{2n})}
\nonumber \\ 
&  &  =
\frac{{\rm Tdet} \left[ \begin{array}{ccc} [g(z_{2 j - 1},z_{2 j};
z_{2k-1},z_{2k}) ]_{j,k=1,\ldots,\alpha} & 
[a(z_{2 j - 1},z_{2 j})]_{j=1,\ldots,\alpha} &   
[h(z_{2 j - 1},z_{2 j};x_k)]_{j=1,\ldots,\alpha; k = 1,\ldots,p} \cr 
[{\hat a}(z_{2 k - 1},z_{2k})]_{k=1,\ldots,\alpha} & \omega & 
[b(x_k)]_{k=1,\ldots,p} \cr 
[{\hat h}(z_{2 k - 1},z_{2k};x_j)]_{j=1,\ldots,p; k=1,\ldots,\alpha} 
& [{\hat b}(x_j)]_{j=1,\ldots,p} & [f(x_j,x_k)]_{j,k = 1,\ldots,p} 
\end{array} \right]}{{\rm Tdet} 
\left[ \begin{array}{cc} [g(z_{2 j - 1},z_{2 j}
;z_{2k-1},z_{2k}) ]_{j,k=1,\ldots,\alpha} & 
[a(z_{2 j - 1},z_{2 j})]_{j=1,\ldots,\alpha} \cr   
[{\hat a}(z_{2 k - 1},z_{2k})]_{k=1,\ldots,\alpha} & \omega  
\end{array} \right]}. \label{rho} 
\end{eqnarray}

Now we make replacement of
the elements of the quaternion kernel
back to those defined in Eq.(\ref{3.13}):
\bea
{\tilde S}(x,y) \rightarrow S(x,y) ,\ \ 
{\tilde D}(x,y) \rightarrow D(x,y) ,\ \ 
{\tilde I}(x,y) \rightarrow I(x,y) ,
\eea
which does not change the values of the quaternion determinants in 
Eqs.(\ref{Xi}) and (\ref{rho}).
We introduce a set of notations 
\begin{eqnarray}
&&S^{II}_{jk}  =  S(z_j,z_k)
\ \ \ \ \ \ (j,k =0,\ldots, 2 \alpha),
\nonumber \\  
&&S^{IR}_{jk}  =  
S(z_j,x_k)
\ \ \ \ \ 
(j = 0,\ldots,2 \alpha;\ 
k = 1,\ldots,p), \\   
&&S^{RR}_{jk}  =  
S(x_j,x_k) 
\ \ \ \ (j,k = 1,\ldots,p), \nonumber
\end{eqnarray} 
and similarly for $D$ and $I$,
and
\begin{eqnarray}
Q^{I}_j & = &  \frac{\Psi_{2 N + 2 \alpha}(z_j)}{s_{2 N + 2 \alpha}}  
\ \ \ \ (j = 0,\ldots,2 \alpha), 
\nonumber \\ 
Q^{R}_j & = &  \frac{\Psi_{2 N + 2 \alpha}(x_j)}{s_{2 N + 2 \alpha}}  
\ \ \ \ (j = 1,\ldots,p), 
\\ 
P^{R}_j & = & \frac{\Psi^{\prime}_{2 N + 2 \alpha}(x_j)}{
s_{2 N + 2 \alpha}}
\ \ \ \ (j = 1,\ldots,p).
\eea
Using Dyson's identity (\ref{Dysoneq}), we can convert the 
quaternion determinant back to a Pfaffian, so that
the correlation functions read
\begin{equation}
\rho(x_1,\ldots,x_p;\{m\}) = 
(-1)^{p(p-1) / 2}
\frac{
{\rm Pf} \left[ \begin{array}{cccc}
-I^{II} & Q^{I} & -I^{IR} & S^{IR} \\ 
- (Q^{I})^{T} & 0 & - (Q^{R})^{T} & - (P^{R})^{T} \\    
(I^{IR})^T & Q^{R} & -I^{RR} & S^{RR} \\ 
-(S^{IR})^{T} & P^{R} & -(S^{RR})^{T} & D^{RR} 
\end{array} \right]  
}{{\rm Pf} \left[ \begin{array}{cc} -I^{II} & Q^{I} \\ 
- (Q^{I})^{T} & 0 \end{array} \right]}. 
\end{equation}
Now we proceed to evaluate 
the component functions in
 the asymptitic limit where
unfolded microscopic variables 
\begin{eqnarray}
\sqrt{2N}z_{0}&\equiv&\z_{0}= 0, \nonumber \\
\sqrt{2N}z_{2 j - 1}&\equiv&
\z_{2j-1}= i \m_j \ \ \ (j=1,\ldots,{\a}), \nonumber \\ 
\sqrt{2N}z_{2 j }&\equiv&
\z_{2j}= -i \m_j \ \ \ \,(j=1,\ldots,{\a}), \\ 
\sqrt{2N}x_{j}&\equiv&
\l_j \ \ \ \ \ \ \ \ \ \ \ \ \ \ \ (j= 1,\ldots,p),
\nonumber
\end{eqnarray}
are kept fixed.
We obtain Eq.(\ref{SDI4}) and
\bml
\bea
&&{\bf Q}_j^I \equiv \frac{1}{\sqrt{2N}} Q_j^I \sim \frac12,\\
&&{\bf Q}_j^R \equiv \frac{1}{\sqrt{2N}} Q_j^R \sim \frac12,\\
&&{\bf P}_j^R \equiv P_j^R \sim 0.
\eea
\eml
\noindent
These matrix elements
constitute
the finite-volume partition function
\begin{eqnarray}
{\cal Z}(\{\m\})
&\equiv &
\Xi_0(\{\frac{\m}{\sqrt{2N}}\})
\nonumber\\
&=&
{\rm const.}\frac{{{\rm Pf} \left[ \begin{array}{cc} 
-{\bf I}^{II} & {\bf Q}^{I} \\ 
- ({\bf Q}^{I})^{T} & 0 \end{array} \right]}}{
\prod_{j=1}^\a \m_j^3 \prod_{j>k}^\a (\m_j^2-\m_k^2)^2},
\label{calZ4odd}
\eea
and 
the scaled spectral correlation functions
\begin{eqnarray}
\rho_s(\l_1,\ldots,\l_p;\{\m\})
&\equiv &
\Bigl(\frac{1}{\sqrt{2N}}\Bigr)^p
\rho(\frac{\l_1}{\sqrt{2N}},\ldots,\frac{\l_p}{\sqrt{2N}};
\{
\frac{\m}{\sqrt{2N}}
\})
\nonumber \\ 
&=& 
(-1)^{p(p-1) / 2}
\frac{
{\rm Pf} \left[ \begin{array}{cccc}
-{\bf I}^{II} & {\bf Q}^{I} & -{\bf I}^{IR} & {\bf S}^{IR} \\ 
- ({\bf Q}^{I})^{T} & 0 & - ({\bf Q}^{R})^{T} & - ({\bf P}^{R})^{T} \\    
({\bf I}^{IR})^T & {\bf Q}^{R} & -{\bf I}^{RR} & {\bf S}^{RR} \\ 
-({\bf S}^{IR})^{T} & {\bf P}^{R} & -({\bf S}^{RR})^{T} & {\bf D}^{RR} 
\end{array} \right]  
}{{\rm Pf} \left[ \begin{array}{cc} 
-{\bf I}^{II} & {\bf Q}^{I} \\ 
- ({\bf Q}^{I})^{T} & 0 \end{array} \right]}. 
\label{rhos4odd}
\end{eqnarray}
It satisfies a sequence
\bea
&&
\rho_s(\{\l\};\mu_1,\mu_1,\ldots,\mu_n,\mu_n,
-\mu_1,-\mu_1,\ldots,-\mu_n,-\mu_n,0,0)
\stackrel{\mu_{n} \to \infty}{\longrightarrow}
\\
&&
\rho_s(\{\l\};\mu_1,\mu_1,\ldots,\mu_{n-1},\mu_{n-1},
-\mu_1,-\mu_1,\ldots,-\mu_{n-1},-\mu_{n-1},0,0)
\stackrel{\mu_{n-1} \to \infty}{\longrightarrow}
\cdots ,
\nonumber
\eea
as each of the masses is decoupled by going to infinity.
To illustrate this decoupling, we exhibit in FIG.4 a plot of
the spectral density $\rho_s(\l;\m,\m,-\m,-\m,0,0)$ 
($p=1, \a=1$).
\Rrule 
\narrowtext

\acknowledgments
This work was supported in part (SMN) by
JSPS Research Fellowships for Young Scientists, and
by Grant-in-Aid No.\ 411044 
from the Ministry of Education, Science, and Culture, Japan.

\appendix
\renewcommand{\theequation}{A.\arabic{equation}}
\setcounter{equation}{0}
\section{quaternion determinant}
A quaternion is defined as a linear combination of 
four basic units $\{1, e_1, e_2, e_3 \}$:
\begin{equation} 
q=q_0+{\bf q} \cdot {\bf e}=q_0+q_1e_1+q_2e_2+q_3e_3. 
\end{equation}
Here the coefficients $q_0, q_1, q_2,$ and $q_3$ are real 
or complex numbers. The first part $q_0$ is called the scalar 
part of $q$. The quaternion basic units satisfy 
the multiplication laws 
\bea 
&& 1 \cdot 1=1,\ \ \ 
1 \cdot e_j=e_j \cdot 1=e_j\ \ (j=1,2,3), 
\nonumber\\
&&e_1^2=e_2^2=e_3^2=e_1e_2e_3=-1. 
\eea
The multiplication is associative and in general not commutative. 
The dual ${\hat q}$ of a quaternion $q$ is defined as 
\begin{equation} 
{\hat q} = q_0-{\bf q} \cdot {\bf e}. 
\end{equation} 
For a selfdual $N \times N$ matrix $Q$
with quaternion elements $q_{jk}$ has a dual 
matrix ${\hat Q}=[{\hat q}_{kj}]$. 
The quaternion units can be represented 
as $2 \times 2$ matrices
\bea 
&&1 \rightarrow 
\left[ \begin{array}{cc} 1 & 0 \\ 0 & 1 \end{array} 
\right], \;\; 
e_1 \rightarrow \left[ \begin{array}{cc} 0 & -1 \\ 1 & 0 \end{array} 
\right],
\nonumber\\
&&
e_2 \rightarrow \left[ \begin{array}{cc} 0 & -i \\ -i & 0 \end{array} \right], 
\;\; 
e_3 \rightarrow \left[ \begin{array}{cc} i & 0 \\ 0 & -i \end{array} \right]. 
\eea
We define a quaternion determinant {\rm Tdet} of a 
selfdual $Q$ (i.e., $Q={\hat Q}$) as
\begin{equation} {\rm Tdet}\, Q = \sum_{P\in S_N}
 (-1)^{N-\ell} \prod_1^\ell (q_{ab} 
q_{bc} \cdots q_{da})_0, 
\end{equation}
where $P$ denotes any permutation of the indices 
$(1,\ldots,N)$ consisting of $\ell$ exclusive cycles of the form 
$(a \rightarrow b \rightarrow c \rightarrow \cdots \rightarrow d \rightarrow a)$ 
and $(-1)^{N-\ell}$ is the parity of $P$. The subscript $0$ means that 
the scalar part of the product is taken over each cycle. Note that a 
quaternion determinant of a selfdual quaternion matrix is always a scalar. 
The quaternion determinant of $Q$ can as well be represented
by its $2 N \times 2 N$ complex matrix representation
$C(Q)$
\cite{DysonQ}:
\begin{equation}
{\rm Tdet}Q = {\rm Pf}[J C(Q)], \ \ \  
J = \openone_N \otimes 
\left[ \begin{array}{cc} 
0 & 1 \\ -1 & 0 \end{array} \right].
\label{Dysoneq}
\end{equation}

\widetext
\newpage
\begin{figure}
\vspace{30mm}
\begin{center}
\leavevmode
\epsfxsize=320pt
\epsfbox{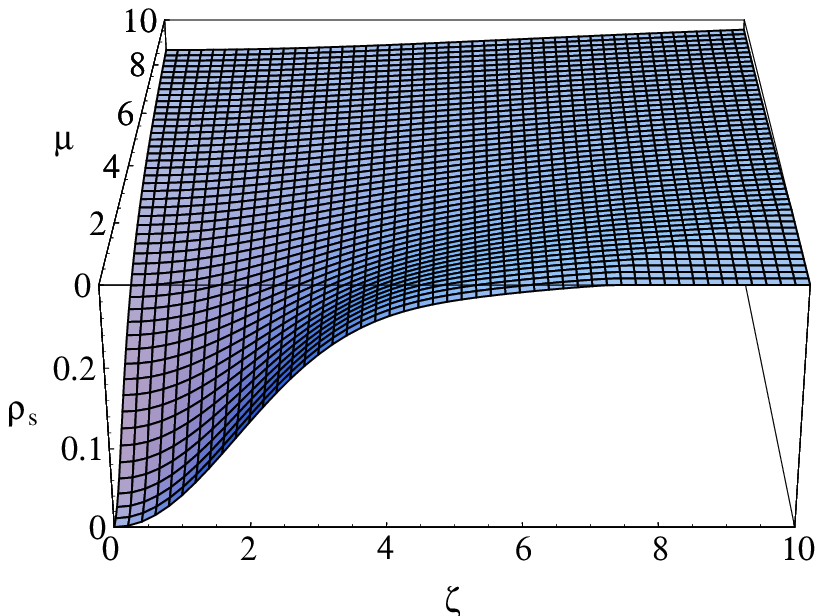}
\end{center}
\caption{The scaled spectral density $\rho_s(\z;\m,-\m)$ for
the orthogonal ensemble with two flavors ($n=1$).}
\vspace{30mm}
\begin{center}
\leavevmode
\epsfxsize=320pt
\epsfbox{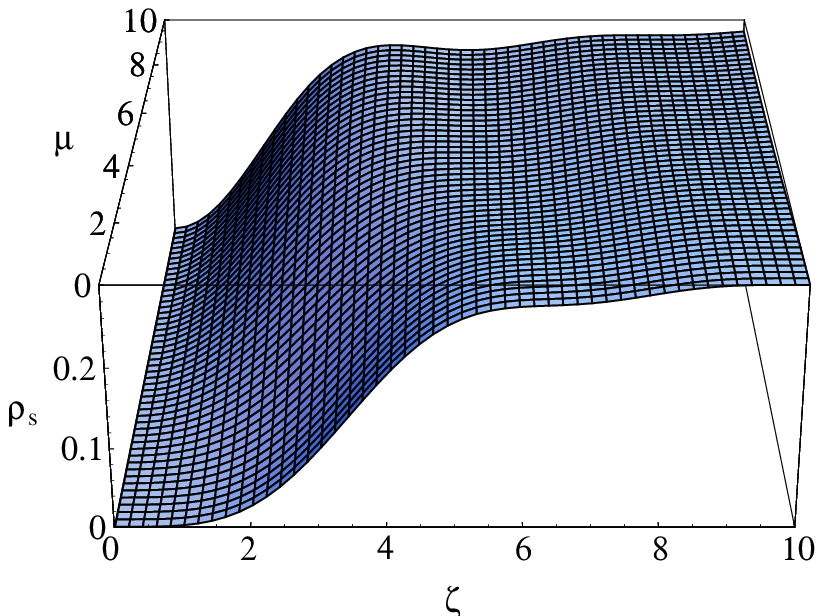}
\end{center}
\caption{The scaled spectral density $\rho_s(\z;\m,-\m,0)$ for
the orthogonal ensemble with three flavors ($n=1$).}
\end{figure}
\newpage
\begin{figure}
\vspace{30mm}
\begin{center}
\leavevmode
\epsfxsize=320pt
\epsfbox{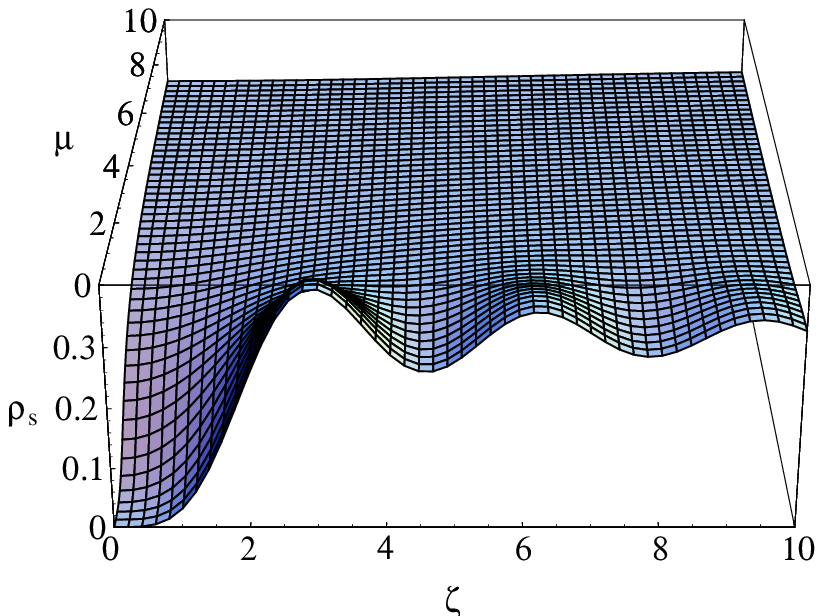}
\end{center}
\caption{The scaled spectral density $\rho_s(\z;\m,\m,-\m,-\m)$ for
the symplectic ensemble with four flavors ($\a=1$).}
\vspace{30mm}
\begin{center}
\leavevmode
\epsfxsize=320pt
\epsfbox{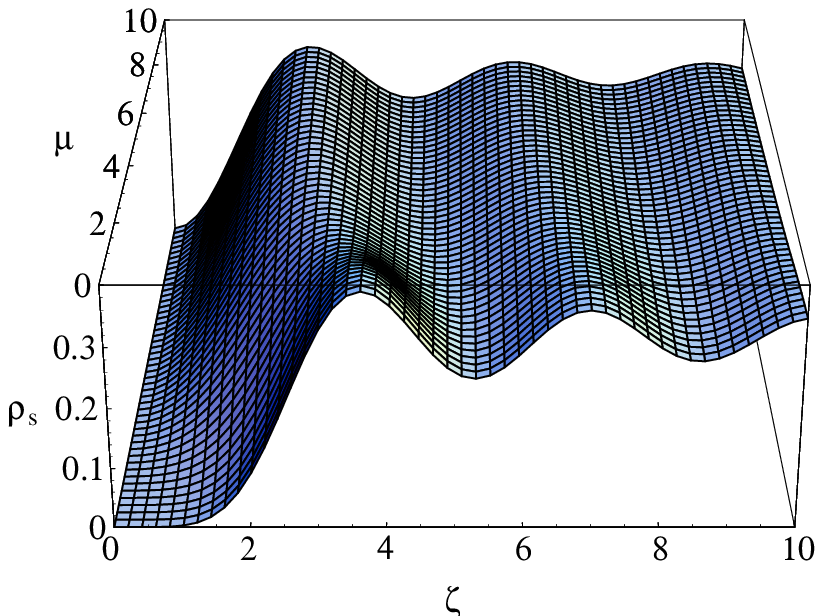}
\end{center}
\caption{The scaled spectral density $\rho_s(\z;\m,\m,-\m,-\m,0,0)$ 
for the symplectic ensemble with six flavors ($\alpha=1$).}
\end{figure}

\begin{references}
\bibitem[\ddagger]{univ}
We tacitly assume that a regularization employed in convoluting
the operators $\varepsilon$ and $K$ in Eq.(\ref{1.21})
does not involve non-universality.
For a detailed discussion, see Ref.\cite{KV}.
\end{references}
\end{document}